\documentclass[]{fairmeta}
\usepackage{amsthm,amsmath,amssymb}
\usepackage{natbib}
\AtBeginDocument{%
  }

\usepackage{enumitem}
\usepackage{multirow}
\usepackage{booktabs}
\usepackage{adjustbox}
\usepackage{etoolbox}
\usepackage{pifont}

\usepackage{graphicx}

\newcommand{\Description}[1]{}
\newcommand{\correspondingmark}{\textsuperscript{\ding{41}}}

\AtBeginEnvironment{tabular}{\small\setlength{\tabcolsep}{4.5pt}}

\begin{document}

\newcommand{\modelname}{PhysOmni}

\title{\modelname{}: Physics-Grounded Multi-Object Scene Generation from a Single Image with Real-Time Interaction}

\author[1,2]{Xin Zhang\textsuperscript{*}\textsuperscript{$\dagger$}}
\author[2]{Yabo Chen\textsuperscript{*}}
\author[2]{Yijie Fang}
\author[1]{Wanying Qu}
\author[2]{Haibin Huang}
\author[2]{Chi Zhang}
\author[1]{Feng Xu\correspondingmark}
\author[2]{Xuelong Li\correspondingmark}

\affiliation[1]{Fudan University}
\affiliation[2]{Institute of Artificial Intelligence (TeleAI), China Telecom}

\contribution{\textsuperscript{*}Equal contributions. \textsuperscript{$\dagger$}Work done during an internship at TeleAI. \correspondingmark Corresponding authors.}
\metadata[Keywords]{Video Generation; Physical Consistency; Scene-level 3D Reconstruction; Controllable Generation; Image-to-Video}
\metadata[Project Page]{\url{https://zephipage.github.io/PhysOmni/}}

\abstract{Recent generative video models achieve impressive visual quality but remain constrained by limited physical consistency and controllability. Existing video generation methods provide minimal physical control, and single-image-to-3D conversion approaches often suffer from object interpenetration. Furthermore, physics-based scene-level 3D generation methods exhibit spatial misalignment, stylized artifacts, and inconsistencies with the input data, restricting their use in realistic interactive video synthesis. We propose \modelname{}, a training-free framework that converts a single image into a physically consistent and controllable video through holistic scene-level 3D reconstruction. By representing the full scene geometry in a unified spatial coordinate system, \modelname{} resolves object penetration and alignment ambiguity. Unlike prior methods, this formulation enables accurate scene-level multi-object interactions and introduces richer, complex control types for advanced mechanics-based manipulation. By decoupling simulation from rendering, \modelname{} bypasses latency-heavy priors, achieving real-time physical interaction previews paired while preserving photorealistic visual fidelity. Experimental results demonstrate that \modelname{} substantially outperforms prior methods in physical fidelity, spatial coherence, and controllability.}

\maketitle

\section{Introduction}

\begin{figure}[t]
  \centering
\includegraphics[width=\textwidth]{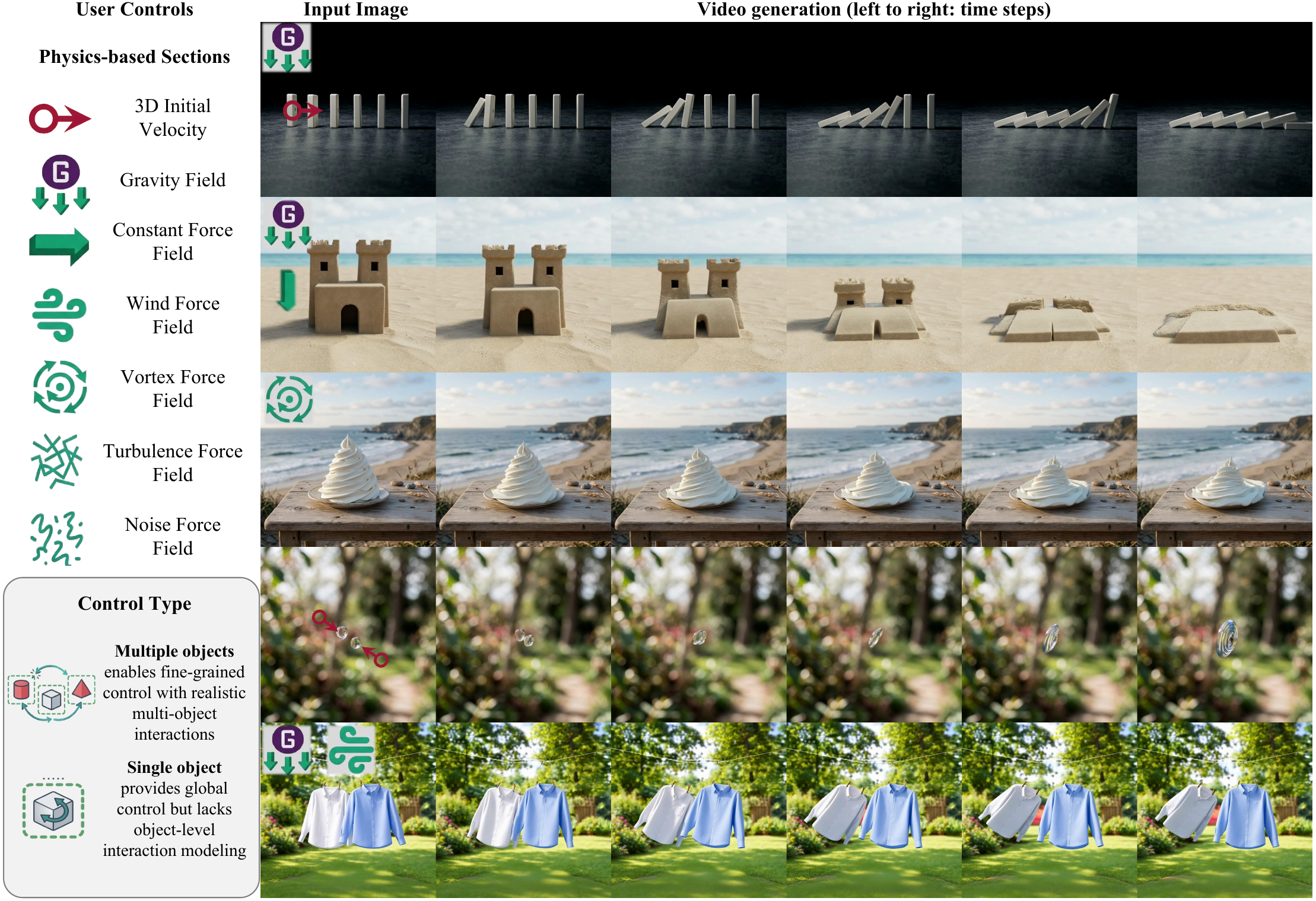}
  \caption{\modelname{} reconstructs a holistic 3D scene from one image for controllable, physics-grounded multi-object video generation. Additional results are provided in Appendix~\ref{sec:additional_results}.}
  \Description{A grid of example results showing single input images on the left and the corresponding generated physically grounded video frames on the right, across several indoor and outdoor scenes with multiple interacting objects.}
\end{figure}

\begin{figure}[t]
    \centering
    \includegraphics[width=\linewidth]{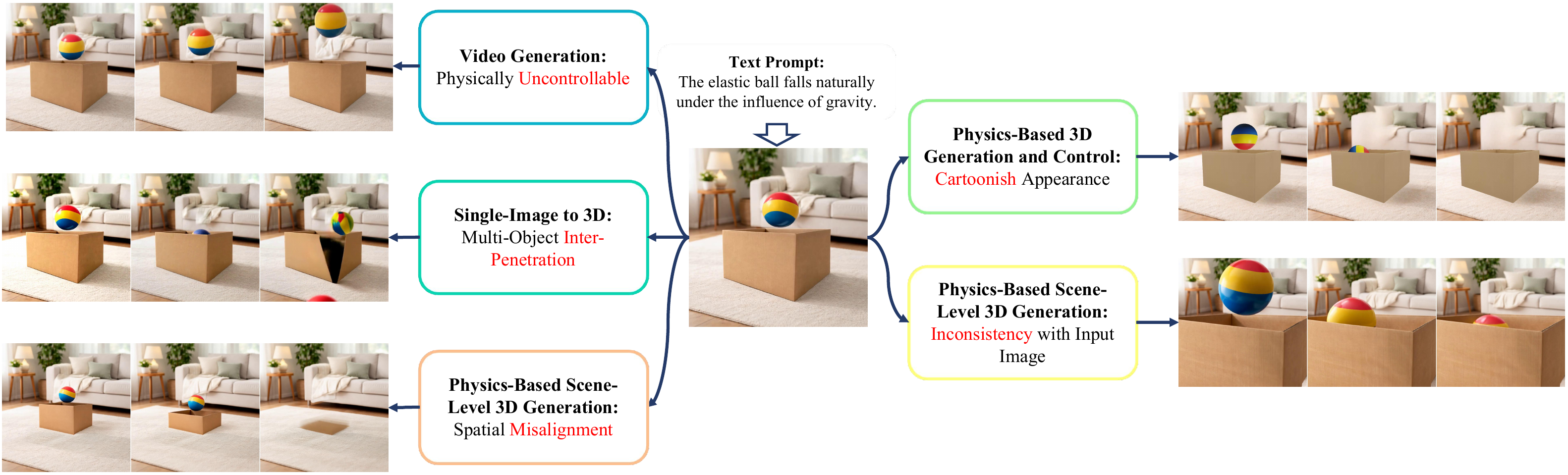}
    \caption{Five limitations of physics-grounded scene generation: (a) uncontrollable motion, (b) object interpenetration, (c) spatial misalignment, (d) cartoonish rendering, and (e) inconsistency with the input image.}
    \label{fig:main_issues}
    \Description{Five side-by-side columns, each showing a failure mode of an existing paradigm for the same falling-ball scenario: uncontrollable video generation, multi-object inter-penetration in single-image-to-3D, spatial misalignment in scene-level 3D generation, cartoonish appearance from 3D generation with control, and inconsistency with the input image.}
\end{figure}

Advances in visual representation and adaptation~\cite{xiang2025denoisingvisiontransformerautoencoder,huang2025diffusiondrivenprogressivetargetmanipulation}, AI-flow systems~\cite{an2026ai,shao2025aiflownetworkedge}, and neural video compression~\cite{chen2026generativevideocompression001,yuan2026enhancingneuralvideocompression} improve visual modeling and delivery, while generative video models~\cite{qualitymacro,huang2025zero,wang2026directingworldfastautoregressive,xiang2026videoweaveunlockinggeometricconsistency,zhang2026symphomotionjointcontrolcamera} achieve strong quality and controllability; however, these developments remain appearance- or systems-driven and lack explicit mechanical reasoning. As summarized in Figure~\ref{fig:main_issues}, generating physics-grounded scenes from a single image is hindered by five primary limitations. First, current video generators are physically uncontrollable (Figure~\ref{fig:main_issues}a) and fail to reliably support object-level interactions or physics-grounded manipulation. Although some studies incorporate physics through auxiliary priors or intermediate signals~\cite{gillman2025force,li2025wonderplay,xie2025physanimator,wang2025physctrl,realwonder2026,zhang2025videorepalearningphysicsvideo,yang2025vlippphysicallyplausiblevideo,satish2026physvideogeneratorphysicallyawarevideo}, they typically rely on manually specified parameters and soft constraints. As a result, they offer limited fine-grained temporal control and yield short, simplistic motions that fail to capture long-horizon, complex physical interactions.

To improve physical reasoning, recent work has explored 3D reconstruction, 4D world modeling~\cite{chen2025teleworlddynamicmultimodalsynthesis,chen2026full4dgeneratingfullscope4d}, and physics-based scene generation from visual inputs. Yet, single-image-to-3D conversion methods~\cite{zhang2024clay,lai2025hunyuan3d25highfidelity3d,xiang2025structured3dlatentsscalable,xu2024instantmesh,wu2025amodal3ramodal3dreconstruction,chen2024cascade,chen2024liftimage3d,li2025triposg} commonly reconstruct objects independently. Despite segmentation-aware formulations~\cite{liu2024part123,chen2024partgen}, this lack of global spatial constraints often causes severe multi-object inter-penetration (Figure~\ref{fig:main_issues}b), while directly combining 3D generation with physical control frequently yields cartoonish appearances (Figure~\ref{fig:main_issues}d). Scene-level 3D generation methods jointly model object layouts and physical plausibility~\cite{paschalidou2021atiss,tang2024diffuscene,yang2024physcene,yang2024scenecraft,zhou2024gala3d,chen2025layout2scene,pat3d2025}, using scalable explicit representations like Gaussian splatting~\cite{zhou2024gala3d,yang2024scenecraft,go2025videorfsplat,lee2024dreamscene360}. Nevertheless, aligning generated scenes with the input image remains challenging. Spatial misalignment (Figure~\ref{fig:main_issues}c) in object pose, scale, and orientation is frequent, and enforcing physical constraints can conflict with image evidence under occlusion and limited viewpoints~\cite{yang2024physcene,yao2025cast,wu2025diorama}, introducing inconsistency with the input image (Figure~\ref{fig:main_issues}e).

We present \modelname{}, a training-free unified framework for holistic 3D scene generation and physics simulation from a single image. Unlike fragmented object-centric pipelines, \modelname{} jointly models all elements within a shared 3D representation under consistent spatial constraints, resolving penetration and alignment ambiguities. This unified formulation enables accurate scene-level multi-object interactions and introduces richer control types for advanced mechanics-based manipulation (e.g., diverse force fields and 3D velocities). By decoupling simulation from rendering, \modelname{} bypasses latency-heavy priors, providing real-time interactive physics previews to transform static images into dynamic, physically coherent environments.

We evaluate \modelname{} on scenes containing single and multiple objects. In single-object settings, it reconstructs geometry from one image and produces stable, physically plausible dynamics under diverse forces, reducing artifacts like drifting and penetration over long-horizon simulations. In challenging multi-object scenes, \modelname{} maintains global spatial coherence and models complex interactions including contact, support, collision, and stacking. Under occlusion and limited viewpoints, the unified representation aligns the pose, scale, and relative layout of objects, mitigating inter-penetration. Quantitative and qualitative results show that \modelname{} outperforms prior methods in physical realism, temporal consistency, and visual fidelity.

Our contributions are summarized as follows:
\begin{itemize}
\item We introduce \modelname{}, a training-free framework enabling accurate scene-level multi-object interactions and complex control types. By decoupling simulation from rendering, it delivers real-time interactive physics previews while maintaining the visual fidelity of the input.
\item We propose a Scene-Aware Pose Alignment mechanism that transforms representations of egocentric meshes into a unified world coordinate system and anchors them to a canonical ground plane, which enforces coherent and physically plausible configurations of the scene.
\item We introduce a coarse-to-fine camera pose optimization strategy for perspective alignment, which ensures accurate photometric and geometric consistency between reconstructed objects and the input image.
\end{itemize}

\section{Related Works}
\label{sec:related_works}

A brief overview of the related literature is provided below. A more comprehensive discussion and an extended review of related works can be found in Appendix~\ref{sec:appendix_related_works}.

\textbf{Controllable and Physics-Aware Video Generation.}
Recent video diffusion and autoregressive models, including reference-controllable long-form generation systems~\cite{huang2026cineweavertrainingfreereferencecontrollablemultishot}, achieve remarkable visual realism but remain predominantly appearance-driven, lacking explicit 3D geometry and an understanding of mechanics.
To bridge this gap, several methods incorporate physical principles through auxiliary priors or intermediate signals~\cite{liu2024physgen,zhang2024physdreamer,xie2025physanimator,gillman2025force,wang2025physctrl,li2025wonderplay,yang2025vlipp,zhang2025videorepalearningphysicsvideo,satish2026physvideogeneratorphysicallyawarevideo,realwonder2026,xiong2026physalign}.
However, these approaches rely on soft constraints and manually specified parameters, which limits their applicability to short and simplistic motions.
In contrast, \modelname{} enables real-time and explicit mechanical control over complex multi-object scenes using diverse force fields (\emph{e.g.}, vortex, wind, and gravity) without relying on latency-heavy priors.

\textbf{Single-Image to 3D Reconstruction.}
Significant advances have been made in lifting single images to 3D representations via SDS-based optimization~\cite{poole2023dreamfusion}, feed-forward networks~\cite{xu2024instantmesh,li2025triposg,zhang2024clay,lai2025hunyuan3d25highfidelity3d,xiang2025structured3dlatentsscalable}, and video diffusion priors~\cite{chen2024liftimage3d,chen2024cascade,wu2025amodal3ramodal3dreconstruction}.
However, these pipelines reconstruct objects in isolation within egocentric coordinate systems.
Even segmentation-aware methods~\cite{liu2024part123,chen2024partgen} lack global spatial awareness, which leads to severe interpenetration when assembling scenes.
\modelname{} introduces a \emph{Scene-Aware Pose Alignment} mechanism that anchors all elements to a canonical ground plane within a unified world coordinate system. This approach inherently resolves ambiguities related to penetration. Because the contact-anchor cue is a generic geometric prior rather than a category-specific one, and material parameters are freely selected by the VLM, \modelname{} generalizes to new scenes and unseen object categories without any category-specific training.

\textbf{Physics-Grounded Scene Generation.}
Recent works jointly model object layouts and physical plausibility by utilizing scalable representations such as Gaussian splatting~\cite{paschalidou2021atiss,tang2024diffuscene,yang2024physcene,yang2024scenecraft,zhou2024gala3d,chen2025layout2scene,lin2025omniphysgs,go2025videorfsplat,lee2024dreamscene360,yao2025cast,wu2025diorama,meng2025scenegen}.
However, the imposition of physical constraints often conflicts with visual evidence under occlusion~\cite{yang2024physcene,yao2025cast,wu2025diorama}, which causes spatial discrepancies or a loss of input fidelity.
\modelname{} addresses this issue through coarse-to-fine camera pose optimization for precise photometric alignment. Crucially, it decouples the physical simulation from the rendering process to deliver real-time interactive physics previews while preserving the original appearance.

\section{Methods}
Existing pipelines for physics-grounded scene generation face several interconnected challenges (Fig.~\ref{fig:main_issues}). Controlling generative video models to enforce physical constraints and target motions remains inherently difficult. Furthermore, current 3D lifting methods often yield physically implausible layouts with intersecting shapes, lack photorealistic details, and tend to drift from the input image, degrading both structural and visual fidelity.

\begin{figure}[t] 
\centering 
\includegraphics[width=\linewidth]{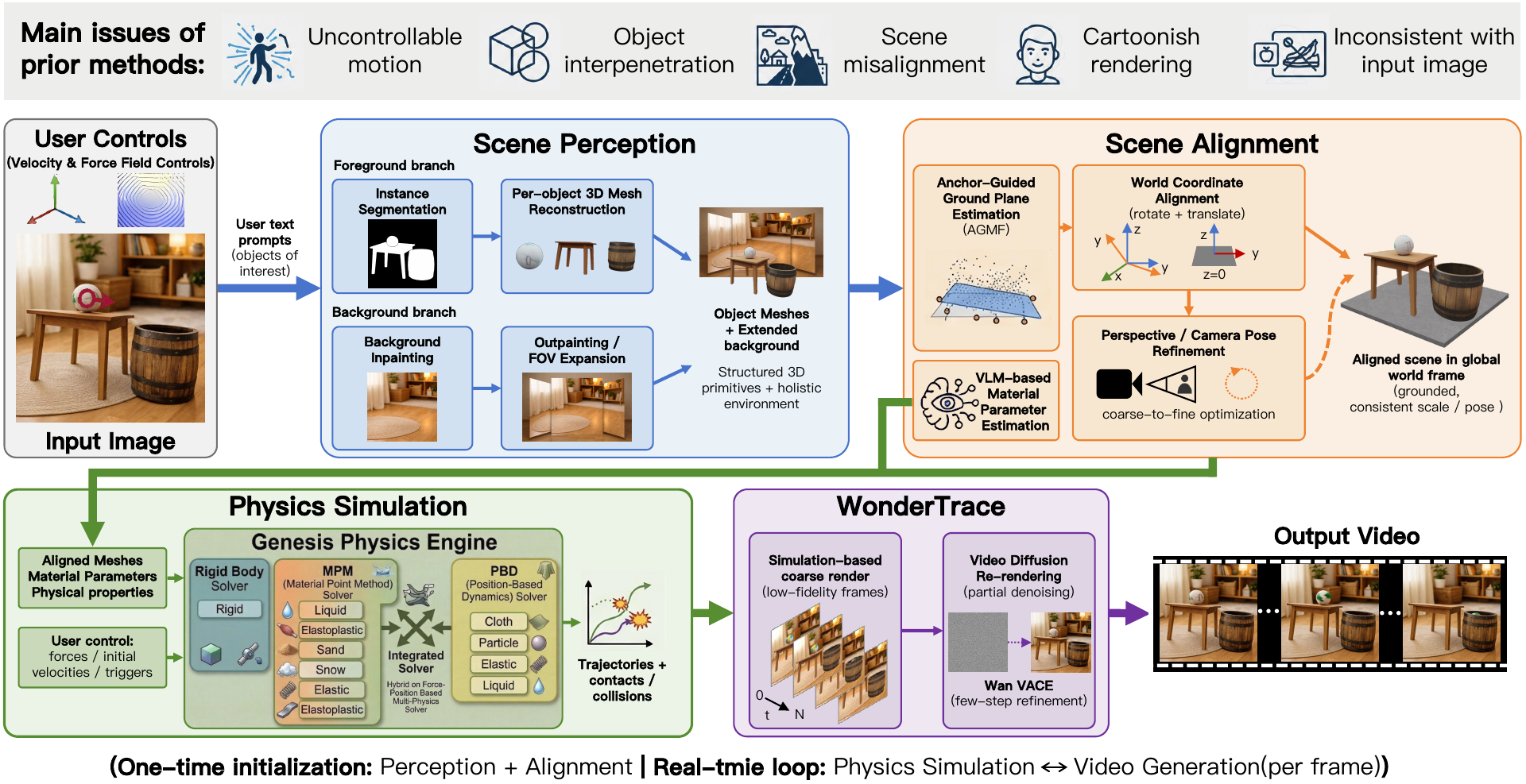} 
\caption{\textbf{Overview of the \modelname{} framework.} \textbf{(a)} Given a single input image and user controls, the pipeline applies Scene Perception to reconstruct 3D object meshes and synthesize a background environment. \textbf{(b)} These components are grounded in a unified global coordinate system through Scene Alignment to ensure geometric consistency, while a VLM-driven parameter estimation module concurrently deduces the physical properties of each entity. \textbf{(c)} Guided by these semantic priors, the Physics Simulation stage computes physically compliant trajectories and collision responses. \textbf{(d)} WonderTrace then bridges the visual domain gap by refining the coarse simulation renders into photorealistic video sequences.}
\Description{A four-stage block diagram flowing left to right: Scene Perception reconstructs object meshes and background from the input image; Scene Alignment places them in a shared world coordinate frame alongside VLM-based physical parameter estimation; Physics Simulation produces trajectories and collision responses; and WonderTrace refines the coarse renders into a photorealistic video.}
\label{fig:pipeline} 
\end{figure}

We address these challenges with \modelname{}, a training-free framework that unifies 3D scene generation and physics-based video synthesis from a single image. The pipeline (Fig.~\ref{fig:pipeline}) operates through four main stages. The Scene Perception module (Section~\ref{sec:scene_perception}) initially decomposes the input into a background and interactive 3D primitives, translating pixel-level data into a manipulable 3D structure. The Scene Alignment stage (Section~\ref{sec:scene_alignment}) then establishes a shared world coordinate frame, assigning consistent poses, scales, and orientations to all entities. This aligns the abstract geometry with the original image perspective to create a physically actionable scene state.

To connect semantic information with physical properties, a Vision-Language Model (VLM) evaluates the material parameters of these aligned entities. The Physics Simulation Engine (Section~\ref{sec:physical_simulation}) processes these assets to resolve multibody dynamics and contact constraints, yielding physically plausible trajectories. To significantly improve rendering speed and bypass the computational cost of traditional graphics rendering while maintaining visual quality, WonderTrace (Section~\ref{sec:video_generation}) refines the simulated frames into photorealistic, temporally coherent sequences using priors from modern video models. Together, these processes convert a static image into an interactive, physically consistent 3D environment that remains faithful to the source and supports long-horizon control.

\subsection{Scene Perception}
\label{sec:scene_perception}
The reconstruction of complex scenes from a single image is an inherently ill-posed problem. To prevent artifacts such as penetration between objects and fragmented geometry, we decompose the input image into instance-level foreground geometry and a globally consistent background(Fig.~\ref{fig:pipeline}a).

\textbf{Reconstruction of Instance-Level Meshes.}
Objects in the foreground are independently segmented using the Segment Anything Model 3~\cite{carion2025sam3segmentconcepts} and reconstructed as explicit 3D meshes via SAM-3D-Objects~\cite{chen2025sam}. Unlike implicit representations, explicit meshes provide well-defined surface topologies, ensuring precise contact interfaces for downstream mechanical simulation and collision handling. These separate meshes are subsequently integrated through coordinate alignment (refer to Appendix~\ref{appendix:scene_perception} for detailed formulations and spatial reasoning).

\textbf{Synthesis and Expansion of Background.}
\label{sec:background_handling}
To eliminate foreground occlusions and ensure spatial consistency, we employ a two-stage background completion strategy. First, the LaMa inpainting model~\cite{suvorov2021resolution} synthesizes occluded regions to yield an unobstructed static view. Second, we use a diffusion-based outpainting model~\cite{fffiloni2025diffusersimageoutpaint} to extend the scene beyond the original field of view. This panoramic context enables physically consistent parallax during camera movement. Detailed implementations of this background synthesis are provided in Appendix~\ref{appendix:scene_perception}.

\subsection{Scene Alignment}
\label{sec:scene_alignment}

\begin{figure}[t]
\centering
\includegraphics[width=\columnwidth]{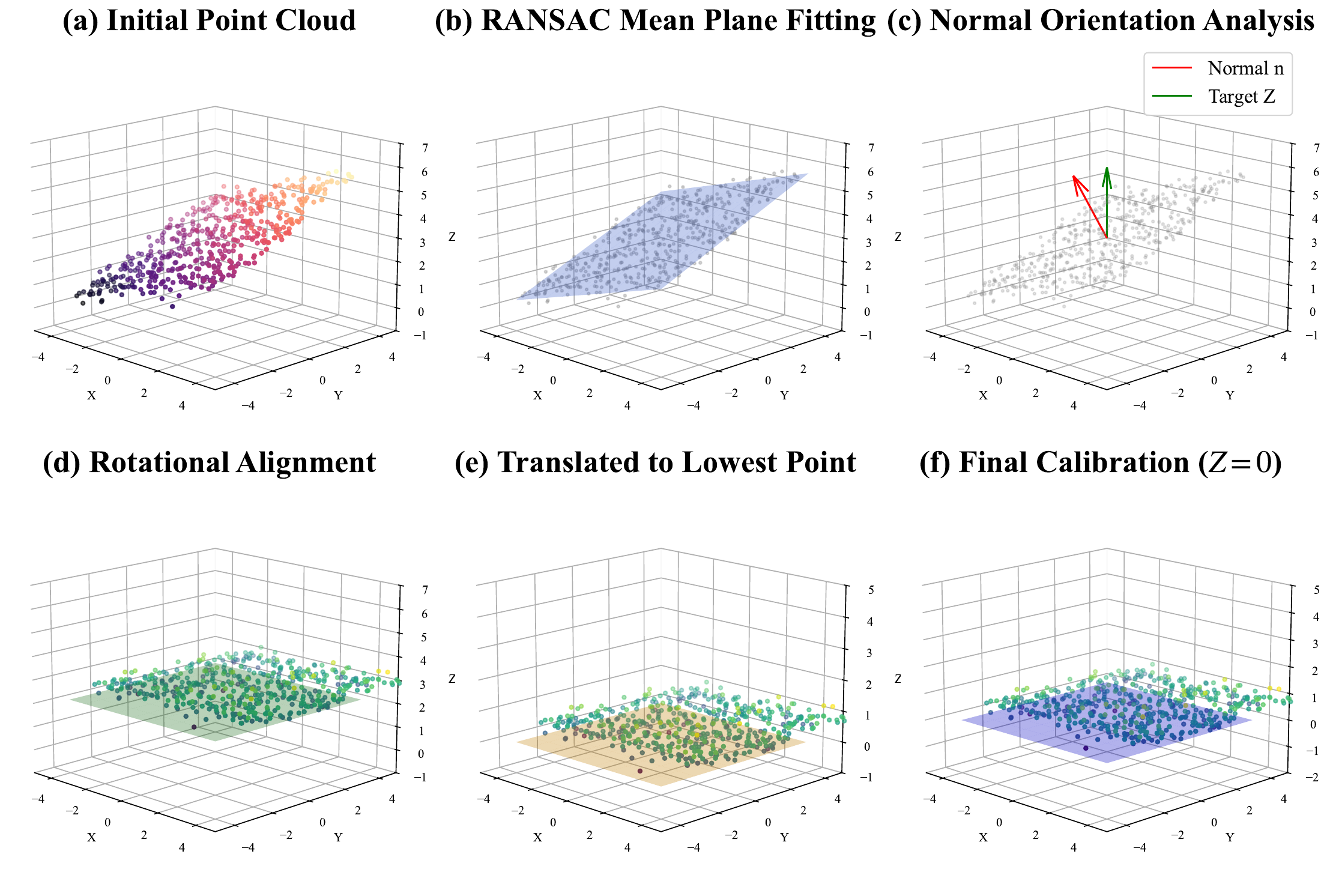}
\caption{\textbf{Ground-plane alignment.} The dominant plane is estimated from the scene point cloud and rigidly transformed to $z=0$ in the gravity-aligned frame $\mathcal{W}$.}
\Description{A diagram showing an aggregated 3D point cloud on the left, the estimated dominant ground plane, and on the right the same scene after a rigid transformation into a gravity-aligned world frame where the ground plane lies at height z equals zero.}
\label{fig:alignment_pipeline}
\end{figure}

\begin{figure}[t]
    \centering
    \includegraphics[width=\columnwidth]{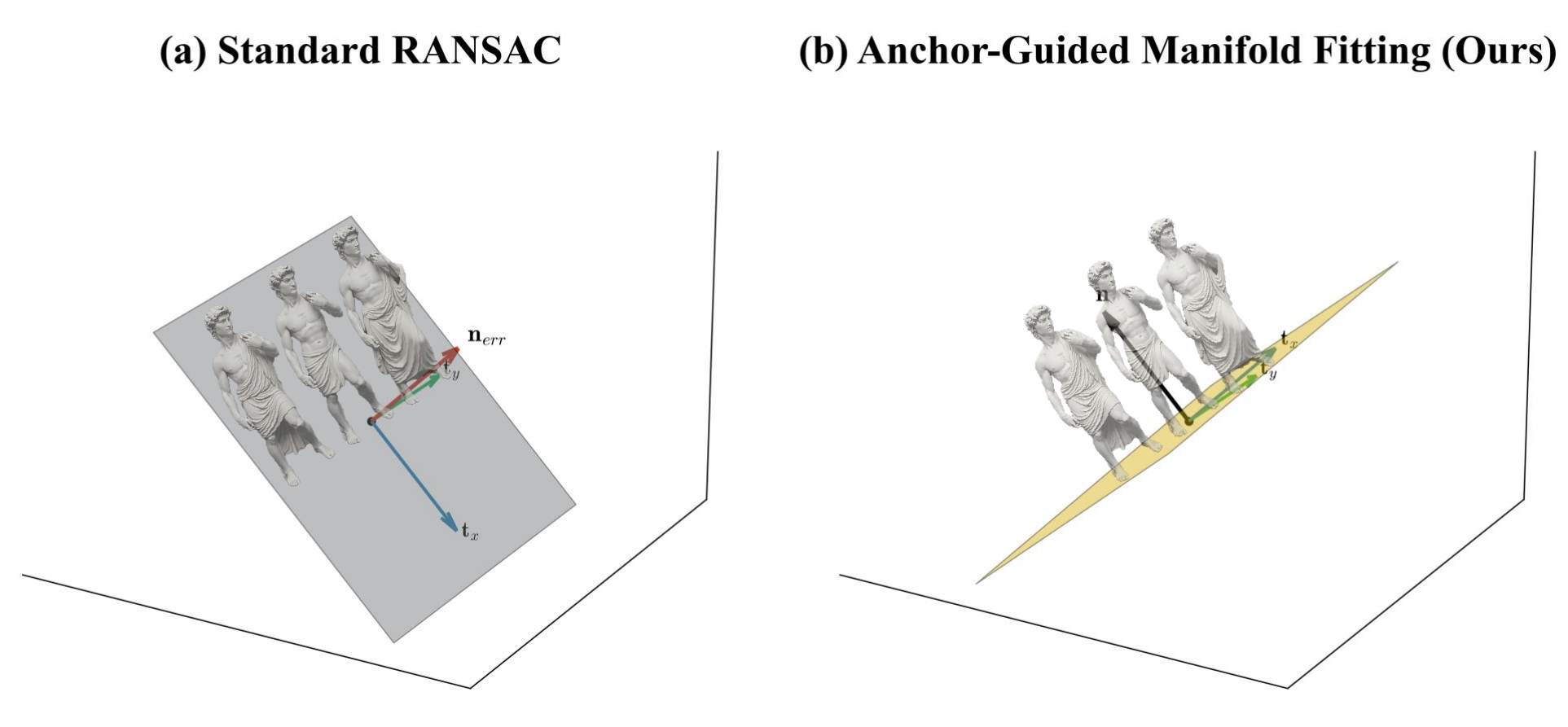}
    \caption{\textbf{Ground-plane estimation.} (a) RANSAC is biased by dense vertical structures. (b) Anchor-Guided Manifold Fitting samples object-base points $\mathcal{P}_{\text{base}}$ to recover the ground normal $\mathbf{n}$.}
    \Description{Two panels comparing ground-plane estimation. Panel (a) shows standard RANSAC fitting a tilted, incorrect plane because vertical structures such as walls and torsos dominate the density. Panel (b) shows the proposed anchor-guided method sampling only low points at object bases, recovering a correct horizontal plane and an accurate upward surface normal.}
    \label{fig:plane_comp}
\end{figure}

To ensure physically valid object interactions and maintain consistency with the input image, we introduce a \textit{Scene Alignment} module(Fig.~\ref{fig:pipeline}b). This module is designed to resolve spatial inconsistencies that stem from egocentric reconstruction and inaccurate camera estimation (Fig.~\ref{fig:main_issues}). It comprises two primary components: \textit{Alignment of Pose Coordinates}, which maps reconstructed meshes into a unified world coordinate system and anchors them to the ground plane, and \textit{Perspective Alignment}, which refines camera parameters to ensure geometric consistency between the reconstructed scene and the input observation.

\subsubsection{Alignment of Pose Coordinates}
Although the previous stage yields detailed mesh representations, these raw descriptors are ill-suited for downstream physical reasoning and interaction tasks. A primary bottleneck arises from the inherent spatial misalignment introduced during the transformation of camera-centric observations into a global world coordinate system (Fig.~\ref{fig:main_issues}). Objects expected to adhere to basic physical constraints often exhibit implausible configurations. For instance, a ball positioned directly above a carton may drift laterally or penetrate the ground plane. To address these spatial misalignments and enforce physical plausibility, we introduce a Scene-Aware Pose Alignment mechanism. This mechanism systematically transforms representations of egocentric meshes into a unified world coordinate frame and anchors them to a canonical ground plane, thereby ensuring coherent spatial alignment and physically valid scene configurations.

To establish a consistent coordinate system for physical interactions, we implement a hierarchical alignment strategy that transforms meshes from the camera coordinate system $\mathcal{C}$ to the world coordinate system $\mathcal{W}$.

\textbf{Canonical Alignment for Single Objects.} For isolated objects, we first perform Global Centroid Normalization. Given a mesh that comprises $N_v$ vertices $V = \{\mathbf{v}_i\}_{i=1}^{N_v}$, the mesh is translated to the origin of the world such that the centroid $\mathbf{\bar{v}}$ satisfies:
\[
\mathbf{\bar{v}} = \frac{1}{N_v} \sum_{i=1}^{N_v} \mathbf{v}_i, \quad \mathbf{v}'_i = \mathbf{v}_i - \mathbf{\bar{v}}.
\]

To determine the principal axes of the object, we apply Principal Component Analysis (PCA) to the zero-centered vertex distribution. The covariance matrix $\mathbf{\Sigma} \in \mathbb{R}^{3 \times 3}$ is computed as follows:
\[
\mathbf{\Sigma} = \frac{1}{N_v} \sum_{i=1}^{N_v} \mathbf{v}'_i (\mathbf{v}'_i)^\top.
\]
Through the eigendecomposition $\mathbf{\Sigma} \mathbf{u}_k = \lambda_k \mathbf{u}_k$, we identify the minor eigenvector $\mathbf{u}_3$ corresponding to the smallest eigenvalue $\lambda_3$. Assuming that most objects rest in a structurally stable configuration where the shortest physical dimension of the object aligns with the direction of gravity, we derive the rotation matrix $\mathbf{R}$ required to align $\mathbf{u}_3$ with the vertical axis of the world coordinate system (Z-up) $[0, 0, 1]^\top$. Finally, a Ground-Contact Correction step is applied:
\[
\mathbf{v}_{\text{final}} = \mathbf{R}\mathbf{v}'_i - [0, 0, Z_{\text{min}}]^\top,
\]
where $Z_{\text{min}}$ denotes the minimum vertical extent of the rotated mesh, which ensures that the object rests precisely on the plane defined by $Z=0$.

\textbf{Scene Alignment for Multiple Objects.} 
Building upon the alignment of individual objects, we extend the approach to complex scenes involving multiple objects. To ensure physical consistency across such multi-object environments, we implement a multi-stage rectification pipeline, as illustrated in Fig.~\ref{fig:alignment_pipeline}. Starting from the constituent meshes of all objects in the scene, denoted as $\mathcal{V}_{\text{all}}$, we first aggregate the vertices to form a global unorganized point cloud $\mathcal{P}$ within the camera frustum. To identify the environmental layout, we employ the RANSAC algorithm to fit a dominant plane to $\mathcal{P}$, defined by the equation $ax + by + cz + d = 0$. The unit normal $\mathbf{n} = [a, b, c]^\top$ thus represents the estimated ground orientation (Fig.~\ref{fig:alignment_pipeline}b-c). To standardize the reference frame for downstream physics-based tasks, we derive an orthogonal transformation $\mathbf{T}_{\text{ext}}$ that aligns the estimated normal $\mathbf{n}$ with the world vertical axis $\mathbf{z}_w = [0, 0, 1]^\top$. Specifically, the rotation $\mathbf{R} \in SO(3)$ is computed via the Rodrigues rotation formula to map $\mathbf{n}$ to $\mathbf{z}_w$, while the translation $\mathbf{t}$ is determined by the minimum vertical coordinate of the rectified points. This transformation effectively anchors the lowest point of the scene to the global ground level (Fig.~\ref{fig:alignment_pipeline}e-f).

\textbf{Estimation of Ground Plane Guided by Anchors.} Standard techniques for plane estimation, such as vanilla RANSAC, often prove suboptimal in spatially complex scenes populated by multiple vertical structures (e.g., standing human subjects). As illustrated in Fig.~\ref{fig:plane_comp}a, the RANSAC algorithm is prone to converging on regions of high point density, such as the torsos of human subjects or walls, rather than the true ground manifold (e.g., the soles of the feet). To address this density-driven bias, we introduce Anchor-Guided Manifold Fitting (AGMF). Unlike global fitting methods that treat all points with equal importance, AGMF shifts the optimization focus toward local geometric extrema that are likely to represent ground-contact points.

The core intuition behind AGMF is to isolate semantic anchors (points physically constrained to the ground) prior to performing the fit. Because the true world vertical is unknown at this stage, we initialize a heuristic gravity vector $\mathbf{g}_{\text{cam}}$ based on standard camera coordinate conventions (e.g., the positive Y-axis corresponding to the visual downward direction). For a scene containing $N$ object meshes, we define an anchor set $\mathcal{P}_{\text{base}}$ by extracting vertices from the lower extrema of each individual mesh $M_i$ along this visual gravity prior. Specifically, we project the vertices onto $\mathbf{g}_{\text{cam}}$ and sample those within a proximity threshold $\delta$ of the maximal projection (i.e., visually lowest) for each object:
\[
\mathcal{P}_{\text{base}} = \bigcup_{i=1}^{N} \left\{ \mathbf{v} \in M_i \mid \mathbf{v}^\top \mathbf{g}_{\text{cam}} \geq \max_{\mathbf{u} \in M_i} (\mathbf{u}^\top \mathbf{g}_{\text{cam}}) - \delta \right\}.
\]

By restricting the input space to $\mathcal{P}_{\text{base}}$, we effectively eliminate the vertical distractors that typically cause the RANSAC method to yield erroneous estimates. We then determine the optimal parameters of the ground plane $\boldsymbol{\pi} = [a, b, c, d]^\top$ by minimizing a robust M-estimator over this refined subset:
\[
\min_{\boldsymbol{\pi}} \sum_{\mathbf{p} \in \mathcal{P}_{\text{base}}} \rho \left( \frac{|a x_p + b y_p + c z_p + d|}{\sqrt{a^2 + b^2 + c^2}} \right),
\]
where $\rho(\cdot)$ denotes a robust loss function (e.g., the Huber or Tukey loss) employed to suppress outliers within the anchor set. This formulation ensures that the resulting surface normal $\mathbf{n} = [a, b, c]^\top$ aligns with the collective gravitational base of all objects in the scene. As demonstrated in Fig.~\ref{fig:plane_comp}b, AGMF yields a physically plausible world orientation, whereas standard RANSAC fails by fitting to dominant vertical geometries.

\subsubsection{Perspective Alignment}
\label{sec:perspective_alignment}

\begin{figure}[t]
\centering
\includegraphics[width=\columnwidth]{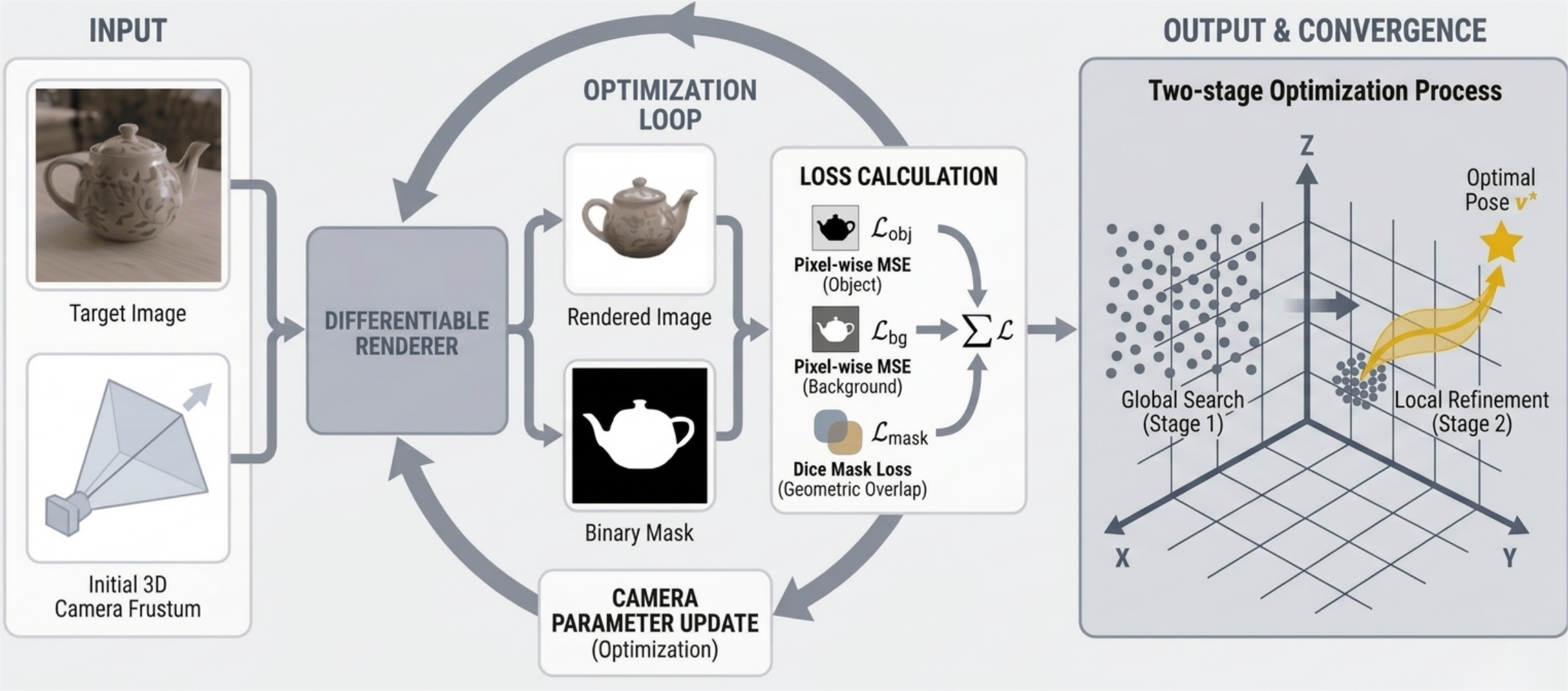}
\caption{Coarse-to-fine camera pose optimization: global stochastic sampling initializes local refinement toward the converged pose $v^*$.}
\label{fig:per_method}
\Description{A schematic of a differentiable rendering loop that optimizes camera pose. The right side visualizes many scattered candidate poses from global stochastic sampling gradually concentrating through local refinement onto a single converged optimal pose labeled v star.}
\end{figure}

As illustrated in Fig.~\ref{fig:main_issues}, naive rendering often results in significant perspective discrepancies—such as inaccurate apparent scale and pose—between the reconstructed objects and the input image. To mitigate these issues, we formulate perspective alignment as a continuous optimization problem over camera parameters to ensure geometric and photometric consistency between the rendered synthetic outputs and the target image (Fig.~\ref{fig:per_method}). Given the non-convex nature of the rendering loss landscape, identifying the global optimum is non-trivial; therefore, we propose a robust coarse-to-fine optimization strategy. The three composite loss terms (an object-region loss, a background loss, and a Dice mask loss) jointly address the classic hard cases of symmetric layouts, textureless surfaces, and foreground--background overlap. This optimized approach effectively eliminates previously observed perspective distortions, yielding renderings that accurately align with the input in spatial configuration, silhouettes, and visual appearance. We refer the reader to Appendix~\ref{sec:appendix_perspective} for the detailed algorithm, including the global stochastic sampling, local derivative-free refinement, and the composite objective functions guiding this process.

\subsection{Physical Simulation}
\label{sec:physical_simulation}

To ensure physically consistent 3D dynamics without relying on unscalable manual assignment, our framework utilizes a Vision-Language Model (VLM) to automatically estimate categorical and continuous material parameters—such as mass, friction, and Young's modulus (detailed in Appendix~\ref{app:vlm_prompt})—for each reconstructed entity based on its visual appearance and scene context. To resolve appearance--material ambiguities (\emph{e.g.}, solid wood versus hollow plastic), we additionally expose a lightweight manual interface that lets users override the per-object material prior before simulation. Following this parameter evaluation, objects are routed to appropriate solvers within a multi-physics backend. While recent physics-based generation methods, such as PhysCtrl~\cite{wang2025physctrl} and PhysGen3D~\cite{chen2025physgen3d}, primarily use the Material Point Method (MPM), this approach can be computationally expensive and may introduce artifacts when handling rigid bodies or stiff constraints. We instead adopt Genesis~\cite{Genesis} as a unified backend(Fig.~\ref{fig:pipeline}c) to address these limitations. Guided by the VLM classifications, the framework decouples material representations across three specialized solvers. Rigid Body Dynamics (RBD) handles non-deformable objects and robotic manipulators; MPM is reserved for fluids and hyperelastic materials undergoing large volumetric deformations; and Position-Based Dynamics (PBD) manages thin-shell structures like cloth and cables. These solvers operate concurrently with GPU-based parallelism and adaptive time-stepping, exchanging force and state information to enable efficient, high-fidelity multi-material interactions. The mathematical formulations and integration mechanisms for these solvers are detailed in Appendix~\ref{app:solvers}.

\subsection{WonderTrace}
\label{sec:video_generation}

To bridge the domain gap between visually coarse physical simulations and photorealistic real-world dynamics, we introduce the WonderTrace module(Fig.~\ref{fig:pipeline}d), which leverages video generation models as a rendering prior to translate simulated frames into high-fidelity temporal sequences. By projecting simulated physical states and trajectories into conditional control signals—such as depth maps, optical flow, and object masks—the module synthesizes photorealistic details while strictly adhering to the engine's underlying structural and temporal dynamics. Furthermore, because the background environment is expanded during the initial Scene Perception stage (Section~\ref{sec:scene_perception}), this rendering process seamlessly supports novel view synthesis and virtual camera movements without out-of-bounds artifacts, with comprehensive implementation details (including specific architectures like Wan2.1 and Wan2.2 VACE) provided in Appendix~\ref{sec:appendix_video_details}.

\section{Experiments}

We evaluate \modelname{} in terms of controllability, physical consistency, visual quality, and runtime efficiency.

\subsection{Experimental Setup}

We evaluate all methods on 60 indoor and outdoor single-image-to-video scenes spanning single- and multi-object interactions, varied object counts, occlusions, and viewpoints. All experiments use one NVIDIA H100 GPU. \modelname{} uses SAM 3~\cite{carion2025sam3segmentconcepts} for segmentation, SAM-3D-Objects~\cite{chen2025sam} for 3D lifting, Genesis~\cite{Genesis} for rigid and deformable simulation, Wan2.1 VACE~\cite{wan2025wan} by default, and Wan2.2 VACE 14B as an optional offline rerenderer. We compare against Sora2-pro~\cite{openaisora22025}, Veo3.1~\cite{deepmind_veo3_2025}, CogVideoX1.5~\cite{yang2024cogvideox}, Wan2.2-A14B~\cite{wan2025wan}, WonderPlay~\cite{li2025wonderplay}, and PhysCtrl~\cite{wang2025physctrl} using the same input image and event description. WonderPlay and PhysCtrl receive only the corresponding official control format with released default settings, while video-prior baselines additionally receive object-level motion text and 2D arrows to provide a comparable control budget. \modelname{} requires approximately 46 s of one-time initialization per scene and then provides an interactive physics preview at approximately 15 FPS (Table~\ref{tab:speed} in the Appendix).

\begin{figure}[t]
  \centering
  \includegraphics[width=\linewidth]{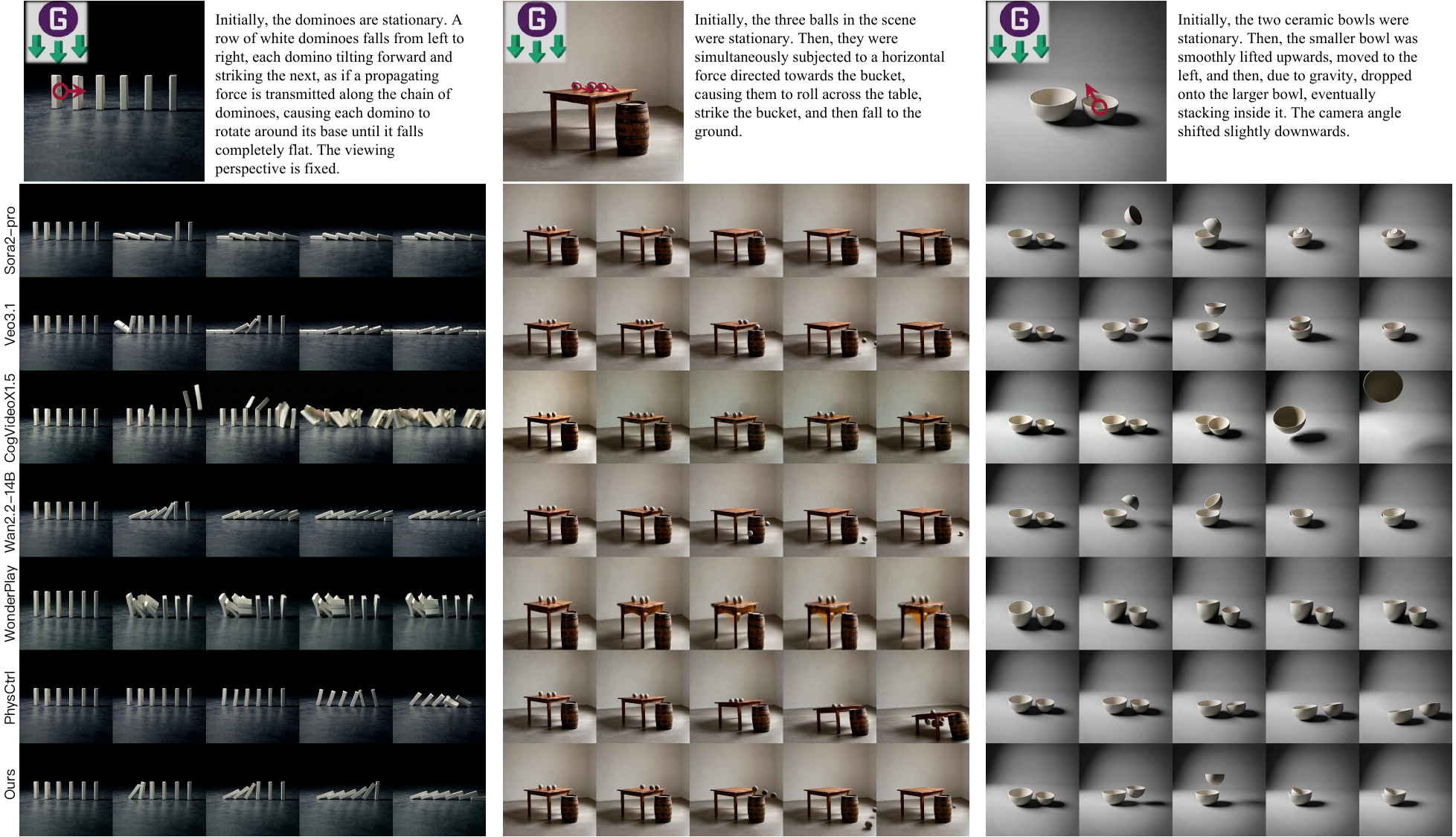}
  \caption{Qualitative comparisons between our approach and existing video generation methods. Additional results are provided in Appendix~\ref{sec:additional_results}.}
  \Description{A comparison grid where each row is a test scene and each column is a method. Our method's frames are shown alongside baselines such as Sora2-pro, Veo3.1, CogVideoX1.5, Wan2.2, WonderPlay, and PhysCtrl, illustrating that our results better preserve object structure and physically plausible motion.}
  \label{fig:compare}
\end{figure}

\subsection{Qualitative Comparison}

Figure~\ref{fig:compare} shows qualitative comparisons with strong video generation baselines. While appearance-driven methods often produce visually appealing motion, they frequently lack reliable control over object trajectories and interactions. In scenes involving multiple objects, these methods tend to generate ambiguous collisions, inconsistent contacts, or motions that deviate from the specified force direction. Although PhysCtrl improves controllability, it still struggles with long-horizon stability and spatial consistency in multi-object settings.

In contrast, \modelname{} produces motions that remain temporally coherent and mechanically consistent with the intended events. The advantage is especially clear in scenarios involving support, collision, and sequential interactions among multiple objects. By reconstructing the scene in a shared world frame and explicitly simulating object dynamics, our method avoids the drift, interpenetration, and identity inconsistency commonly observed in video-prior baselines.

\subsection{Quantitative Comparison}

We quantitatively evaluate the 60-scene test set for controllability, physical plausibility, and overall video quality. Following the VideoPhy~\cite{bansal2024videophy} protocol, we adopt a GPT-5-based 5-point Likert scoring framework. For each method, we evaluate 60 generated videos under three complementary criteria. Semantic adherence(SA) measures alignment between the generated video and text prompt, emphasizing whether specified initial forces or velocities and resulting motions match intended dynamics. Physical commonsense(PC) assesses whether generated motions and interactions are plausible under basic physical principles like gravity, inertia, and contact constraints. Video quality(VQ) evaluates visual fidelity, frame-level realism, and temporal smoothness.

Table~\ref{tab:results1} reports the GPT-5 evaluation results. Our method achieves the best performance on semantic adherence and physical commonsense, showing that explicit scene reconstruction and simulator-based dynamics substantially improve controllability and physical plausibility. Notably, the lower scores of PhysCtrl and WonderPlay stem from our test set predominantly featuring multi-object scenes, where earlier methods struggle to accurately model complex physical interactions. Although Veo3.1 attains the highest video quality score, \modelname{} remains competitive while providing substantially stronger physical grounding and interaction correctness. These results support the core design of \modelname{}: decoupling physical reasoning from appearance synthesis leads to better control without sacrificing perceptual quality.

\begin{table}[t]
\caption{Quantitative comparison}
\label{tab:results1}
\begin{center}
\begin{tabular}{lccc}
\toprule
\textbf{Method} & \textbf{SA$\uparrow$} & \textbf{PC$\uparrow$} & \textbf{VQ$\uparrow$} \\
\midrule
PhysCtrl~\cite{wang2025physctrl}      & 1.85 & 2.34 & 2.58 \\
WonderPlay~\cite{li2025wonderplay}    & 2.24 & 2.49 & 2.97 \\
CogVideoX1.5~\cite{yang2024cogvideox} & 2.36 & 2.02 & 2.42 \\
Wan2.2-A14B~\cite{wan2025wan}         & 2.46 & 2.68 & 3.46 \\
Sora2-pro~\cite{openaisora22025}      & 2.82 & 2.35 & 3.58 \\
Veo3.1~\cite{deepmind_veo3_2025}      & 2.66 & 2.92 & \textbf{3.71} \\
\textbf{Ours}                         & \textbf{3.29} & \textbf{3.15} & 3.61 \\
\bottomrule
\end{tabular}
\end{center}
\footnotesize
SA: Semantic Adherence. PC: Physical Commonsense. VQ: Video Quality.
\end{table}

\subsection{Human Evaluation}

Participants rank seven videos per scene by semantic adherence, physical commonsense, video quality, and overall preference, and we aggregate the rankings using the Borda score $s_i=8-\pi_i$.

Table~\ref{tab:results2} shows that \modelname{} ranks first on all four criteria, with particularly large margins in semantic adherence and physical commonsense, indicating improved controllability, physical plausibility, and overall preference.

\begin{table}[t]
\caption{Human evaluation}
\label{tab:results2}
\begin{center}
\begin{tabular}{lcccc}
\toprule
\textbf{Method}
& \textbf{SA$\uparrow$}
& \textbf{PC$\uparrow$}
& \textbf{VQ$\uparrow$}
& \textbf{UP$\uparrow$} \\
\midrule
PhysCtrl~\cite{wang2025physctrl}      & 3.45 & 3.12 & 3.29 & 3.31 \\
WonderPlay~\cite{li2025wonderplay}    & 2.02 & 2.05 & 1.98 & 2.06 \\
CogVideoX1.5~\cite{yang2024cogvideox} & 2.78 & 2.81 & 2.86 & 2.84 \\
Wan2.2-A14B~\cite{wan2025wan}         & 3.72 & 3.85 & 3.73 & 3.76 \\
Sora2-pro~\cite{openaisora22025}      & 4.38 & 4.43 & 4.67 & 4.28 \\
Veo3.1~\cite{deepmind_veo3_2025}      & 4.72 & 4.84 & 4.83 & 4.84 \\
\textbf{Ours}                         & \textbf{6.93} & \textbf{6.90} & \textbf{6.64} & \textbf{6.91} \\
\bottomrule
\end{tabular}
\end{center}
\footnotesize
Participants rank seven videos per scene, one per method. Scores are aggregated using Borda count ($8-\mathrm{rank}$), yielding a score range of $[1,7]$.
\end{table}

\subsection{Ablation Study}

We perform ablation studies to isolate the contribution of key components in \modelname{}, including scene alignment, camera optimization, and simulation-aware initialization. Additional results and analyses are provided in Appendix~\ref{appendix:ablation_study}.

\begin{table}[htbp]
\caption{Ablation on scene-aware pose alignment.}
\label{tab:ablation_alignment}
\centering
\begin{adjustbox}{max width=\linewidth,center}
\begin{tabular}{lcccc}
\toprule
\textbf{Variant} & \textbf{Mask IoU$\uparrow$} & \textbf{PR$\downarrow$} & \textbf{SVR$\downarrow$} & \textbf{ISR$\uparrow$} \\
\midrule
w/o alignment & 0.11 & \textbf{0.00} & 1.00 & 0.00 \\
+ single-obj. canonical alignment & 0.35 & 0.50 & 0.15 & 0.88 \\
+ vanilla RANSAC plane fitting & \textbf{0.54} & 0.01 & 0.15 & 0.88 \\
+ AGMF (full pose alignment) & 0.54 & \textbf{0.00} & 0.15 & \textbf{0.90} \\
\bottomrule
\end{tabular}
\end{adjustbox}
\parbox{\linewidth}{\footnotesize PR: Penetration Rate. SVR: Support Violation Rate. ISR: Interaction Success Rate.}
\end{table}

\textbf{Scene-aware pose alignment.} Table~\ref{tab:ablation_alignment} shows that removing alignment yields poor grounding (Mask IoU 0.11), complete support failure (SVR 1.00), and no successful interactions (ISR 0.00); its zero penetration is therefore trivial. Canonical single-object alignment improves Mask IoU to 0.35 but raises PR to 0.50, while RANSAC reaches a Mask IoU of 0.54 and reduces PR to 0.01. AGMF retains this spatial accuracy, eliminates penetration, and achieves the highest ISR of 0.90, demonstrating robust physically valid initialization.

\begin{table}[htbp]
\caption{Ablation on perspective alignment.}
\label{tab:ablation_camera}
\centering
\begin{adjustbox}{max width=\linewidth,center}
\begin{tabular}{lcccc}
\toprule
\textbf{Variant} & \textbf{SSIM$\uparrow$} & \textbf{LPIPS$\downarrow$} & \textbf{Mask IoU$\uparrow$} & \textbf{Reproj. Err.$\downarrow$} \\
\midrule
Initial camera only & 0.76 & 0.14 & 0.26 & 38.95 \\
Global search only & 0.76 & 0.13 & 0.44 & 20.97 \\
Local Powell only & \textbf{0.77} & \textbf{0.13} & \textbf{0.55} & 14.77 \\
Coarse-to-fine (ours) & 0.76 & 0.13 & 0.54 & \textbf{14.33} \\
\bottomrule
\end{tabular}
\end{adjustbox}
\parbox{\linewidth}{\footnotesize Reproj. Err. denotes the average reprojection error in pixels after optimization.}
\end{table}

\textbf{Perspective alignment.} Table~\ref{tab:ablation_camera} shows that the initial camera has a 38.95-pixel reprojection error and 0.26 Mask IoU. Global search and local Powell optimization reduce the error to 20.97 and 14.77 pixels, respectively, while our coarse-to-fine strategy achieves the lowest error of 14.33 pixels with comparable visual quality (SSIM 0.76; LPIPS 0.13) and Mask IoU (0.54). This confirms the benefit of combining global exploration with local refinement.

\section{Conclusion}

We present \modelname{}, a training-free framework that converts a single image into physically controllable video through holistic 3D reconstruction, scene-aware pose and perspective alignment, explicit simulation, and photorealistic rerendering. Experiments demonstrate improved multi-object interaction, spatial coherence, and long-horizon control, while ablations validate the alignment and camera optimization. Performance remains limited by monocular segmentation and reconstruction errors, particularly under severe occlusion and for fluids or thin shells.

\bibliography{paper}
\bibliographystyle{authordate1}

\clearpage

\appendix

\begin{center}
    \vspace*{1em}
    {\Huge \bfseries \modelname{}: Physics-Grounded Multi-Object Scene Generation from a Single Image with Real-Time Interaction\par}
    {\Large Supplementary Material\par}
    \vspace{1em}
\end{center}

\section{Extended Related Work}
\label{sec:appendix_related_works}

\subsection{Controllable and Physics-Aware Video Generation}
\label{sec:app_physics_video}

Recent generative video models based on diffusion~\cite{ho2022video} and autoregressive architectures achieve unprecedented visual realism in the synthesis of dynamic content.
However, these models are predominantly appearance-driven and lack an explicit understanding of three-dimensional geometry and mechanical laws.
To address this limitation, several methods attempt to incorporate physics through auxiliary priors, drag-based controls, or intermediate physical signals.
PhysGen~\cite{liu2024physgen} integrates rigid-body simulation with video diffusion to generate physically plausible image-to-video content conditioned on applied forces.
PhysDreamer~\cite{zhang2024physdreamer} endows static 3D Gaussian objects with interactive dynamics by leveraging video generation priors for the estimation of material properties.
PhysAnimator~\cite{xie2025physanimator} combines deformable-body simulation with sketch-guided video diffusion for physically grounded cartoon animation.
Force Prompting~\cite{gillman2025force} finetunes video models on simulator-synthesized data to follow localized point forces and global wind fields.
PhysCtrl~\cite{wang2025physctrl} learns a generative physics network that produces 3D point trajectories conditioned on physical parameters across multiple material types.
WonderPlay~\cite{li2025wonderplay} introduces a hybrid generative simulator that couples a physics solver with a video generator for action-conditioned dynamic 3D scene synthesis.
VLIPP~\cite{yang2025vlipp} proposes a two-stage framework that first predicts coarse physical trajectories via vision-language reasoning, and subsequently conditions a controllable diffusion model on the resulting optical flow.
VideoREPA~\cite{zhang2025videorepalearningphysicsvideo} distills physics understanding from self-supervised video encoders into text-to-video diffusion models through token-level relational alignment.
PhysVideoGenerator~\cite{satish2026physvideogeneratorphysicallyawarevideo} embeds a learnable physics prior by aligning diffusion latents with V-JEPA~2 representations via cross-attention.
RealWonder~\cite{realwonder2026} achieves real-time action-conditioned generation at 13.2 frames per second by bridging single-image 3D reconstruction, physics simulation, and a distilled four-step video generator.
PhysAlign~\cite{xiong2026physalign} constructs a fully controllable synthetic data pipeline based on rigid-body simulation and proposes a unified physical latent space that couples 3D geometry constraints with Gram-based spatio-temporal relational alignment.
More recently, large-scale video foundation models for Physical AI, such as Cosmos~\cite{nvidia2026worldsimulationvideofoundation}, learn world simulation directly from massive video corpora and demonstrate strong general-purpose dynamics priors.
These world models, however, primarily target learned, implicit dynamics for data generation and policy learning, and do not expose explicit, interactive control over per-object forces or provide the deterministic, real-time physical previews that \modelname{} offers.

Despite these advances, existing approaches rely heavily on soft constraints and manually specified parameters.
This reliance limits the respective methods to short, simplistic motions that fail to support fine-grained, long-horizon physical interactions.
In contrast, \modelname{} enables real-time, explicit mechanical control over complex multi-object scenes using diverse force fields (\emph{e.g.}, vortex, wind, gravity) without relying on latency-heavy priors.

\subsection{Single-Image to 3D Reconstruction}
\label{sec:app_image_to_3d}

Significant progress has been made in the conversion of single images to 3D models.
Score Distillation Sampling (SDS)~\cite{poole2023dreamfusion} pioneered the lifting of 2D diffusion priors into 3D optimization, inspiring a family of methods that employ triplanes, large-scale feed-forward networks, and rectified flow transformers.
CLAY~\cite{zhang2024clay} trains a 1.5B-parameter latent diffusion transformer on diverse 3D geometries with multi-resolution VAE encoding.
Hunyuan3D~2.5~\cite{lai2025hunyuan3d25highfidelity3d} scales a shape foundation model (LATTICE) to 10B parameters for the generation of sharp, detailed meshes with support for physically based rendering textures.
TRELLIS~\cite{xiang2025structured3dlatentsscalable} introduces Structured 3D Latents (SLAT) decoded into radiance fields, Gaussians, or meshes via 2B-parameter rectified flow transformers trained on 500K objects.
InstantMesh~\cite{xu2024instantmesh} combines multi-view diffusion with a transformer-based sparse-view reconstruction model for the generation of meshes within 10 seconds.
Amodal3R~\cite{wu2025amodal3ramodal3dreconstruction} addresses occluded objects through mask-weighted cross-attention and occlusion-aware conditioning for amodal 3D reconstruction.
Cascade-Zero123~\cite{chen2024cascade} proposes a self-prompted cascade framework that progressively generates nearby views to improve multi-view consistency.
LiftImage3D~\cite{chen2024liftimage3d} leverages latent video diffusion models with articulated camera trajectories and distortion-aware 3D Gaussian splatting.
TripoSG~\cite{li2025triposg} employs large-scale rectified flow models with an advanced SDF-based VAE to achieve state-of-the-art shape fidelity.

However, these pipelines typically reconstruct individual objects in isolation within egocentric coordinate systems.
Even with segmentation-aware formulations such as Part123~\cite{liu2024part123} and PartGen~\cite{chen2024partgen}, the methods lack global spatial awareness.
When independently reconstructed objects are composed into a single scene, severe spatial misalignment and interpenetration inevitably arise.
Unlike these fragmented object-centric methods, \modelname{} introduces a \emph{Scene-Aware Pose Alignment} mechanism that anchors all elements to a canonical ground plane and transforms the elements into a unified world coordinate system.
This approach inherently resolves penetration ambiguities and guarantees geometrically coherent configurations.

\subsection{Physics-Grounded Scene Generation and Simulation}
\label{sec:app_scene_gen}

Recent literature explores physics-based scene-level generation, jointly modeling object layouts and physical plausibility.
ATISS~\cite{paschalidou2021atiss} formulates indoor scene synthesis as the autoregressive generation of object attributes via transformers.
DiffuScene~\cite{tang2024diffuscene} extends this approach with denoising diffusion models operating on unordered object sets, synthesizing physically plausible indoor scenes.
PhyScene~\cite{yang2024physcene} incorporates physical interaction constraints into 3D scene synthesis for embodied AI.
SceneCraft~\cite{yang2024scenecraft} leverages layout-guided diffusion with 3D bounding boxes for the generation of complex indoor scenes.
GALA3D~\cite{zhou2024gala3d} proposes layout-guided generative Gaussian splatting with priors generated by LLMs and compositional diffusion optimization.
Layout2Scene~\cite{chen2025layout2scene} employs semantic-guided geometry and appearance diffusion priors conditioned on 3D semantic layouts.
PAT3D~\cite{lin2025omniphysgs} integrates differentiable rigid-body contact simulation into the text-to-3D pipeline, introducing simulation-in-the-loop optimization for intersection-free, physically stable scenes.
VideoRFSplat~\cite{go2025videorfsplat} jointly models multi-view images and camera poses via a dual-stream architecture for direct text-to-3D Gaussian generation.
DreamScene360~\cite{lee2024dreamscene360} generates 360-degree scenes by lifting panoramic images into 3D Gaussians with a globally optimized point cloud initialization.
CAST~\cite{yao2025cast} reconstructs component-aligned 3D scenes from single RGB images using physics-aware correction with fine-grained relation graphs, and was nominated for the Best Paper award at SIGGRAPH 2025.
Diorama~\cite{wu2025diorama} proposes a zero-shot pipeline that decomposes single-view scene modeling into perception and CAD-based assembly with semantic-aware layout optimization.
SceneGen~\cite{meng2025scenegen} generates multiple 3D assets from a scene image in a single feedforward pass with local-global feature aggregation.

However, accurately aligning the generated scenes with the original input image remains a critical challenge.
Imposing physical constraints often conflicts with visual evidence under occlusion or limited viewpoints~\cite{yang2024physcene,yao2025cast,wu2025diorama}, which results in spatial discrepancies, stylized artifacts, or a loss of input fidelity.
To overcome this challenge, \modelname{} employs a coarse-to-fine camera pose optimization strategy to ensure precise photometric and geometric alignment.
Crucially, by decoupling the underlying physical simulation from the rendering pipeline, the proposed framework uniquely bypasses the latency bottlenecks of previous methods.
This decoupling delivers real-time interactive physics previews while perfectly preserving the high-fidelity appearance of the original input.

\section{Details of Scene Perception}
\label{appendix:scene_perception}
This section provides the implementation details, mathematical formulations, and detailed justifications for the Scene Perception module introduced in the main paper.

\textbf{Instance-level Mesh Reconstruction.}
To prevent geometric interference, objects are processed within local coordinate frames. Given an input image $I \in \mathbb{R}^{H \times W \times 3}$ and a set of semantic prompts $P$, we employ the Segment Anything Model 3 (SAM 3)~\cite{carion2025sam3segmentconcepts} to generate high-fidelity instance masks $\{M_i\}_{i=1}^N$. These masks are then utilized by SAM-3D-Objects~\cite{chen2025sam} to lift the 2D representations into 3D geometric primitives, outputting explicit meshes $\mathcal{S}_i$. The joint processing of instance masks and image features at this stage is crucial, as it preserves the relative spatial configuration and scale of individual entities.

We explicitly avoid implicit or point-based representations because they are prone to ghosting artifacts and numerical instability during contact resolution. Explicit meshes allow for the accurate computation of collision manifolds and friction forces, thereby satisfying fundamental physical constraints during mechanical simulation.

\textbf{Background Synthesis and Expansion.}
To accommodate dynamic camera viewpoints, resolve scale ambiguity, and mitigate the "cardboard effect" common in localized reconstruction methods, we reconstruct the background sequentially. 

For the first stage, we synthesize the occluded regions using the LaMa inpainting model~\cite{suvorov2021resolution} to derive a restored background $B_{\text{rest}}$:
\[
B_{\text{rest}} = \mathcal{F}_{\text{inpaint}}\left(I, \bigcup_{i=1}^N M_i\right)
\]
This operation effectively removes the foreground masks $\{M_i\}_{i=1}^N$ from the input image $I$, yielding a clean canvas.

In the second stage, we expand the field of view using a diffusion-based outpainting model~\cite{fffiloni2025diffusersimageoutpaint}:
\[
B_{\text{ext}} = \mathcal{F}_{\text{outpaint}}(B_{\text{rest}}, \Phi)
\]
where $\Phi$ denotes the target aspect ratio and structural expansion parameters. The resulting extended background $B_{\text{ext}}$ provides necessary constraints on vanishing points and global layout, enabling consistent foreground-background parallax.

\section{Details of Perspective Alignment}
\label{sec:appendix_perspective}

\begin{figure}[!t]
    \centering
    \includegraphics[width=\linewidth]{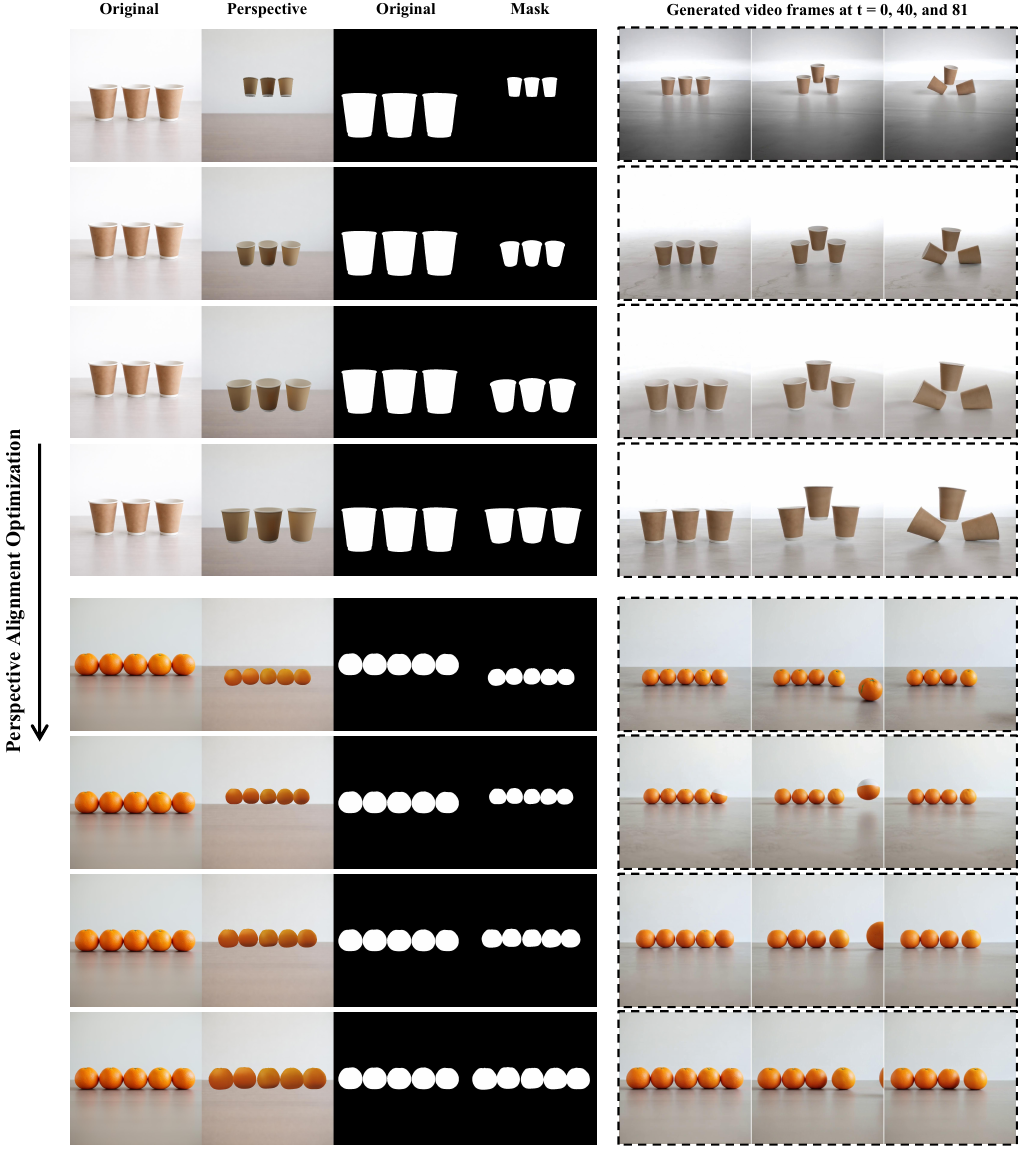}
    \caption{\textbf{Qualitative visualization of the perspective alignment optimization process.} From top to bottom, the rendering loss decreases gradually as the camera parameters undergo refinement. Each row displays the original input image, the rendered perspective under the current camera parameters, and the corresponding binary mask.}
    \Description{Several rows arranged top to bottom showing successive iterations of camera optimization. In each row, from left to right, are the input image, the rendered object under the current camera parameters, and its binary mask; lower rows show progressively tighter alignment between the render and the input.}
    \label{fig:per_overview}
\end{figure}

This section provides the detailed mathematical formulations and procedural steps for the perspective alignment optimization introduced in the main paper. The success of this coarse-to-fine approach in eliminating perspective distortions and aligning spatial configurations is comprehensively demonstrated in Fig.~\ref{fig:per_overview}.

\textbf{Coarse-to-Fine Optimization Strategy.} Initially, we conduct a coarse global search by uniformly sampling candidate poses within a bounded space to identify a reliable initialization, thereby mitigating the risk of convergence to poor local minima. Subsequently, we execute a local refinement step using the derivative-free Powell optimization method under box constraints. This approach enables precise pose adjustment without reliance on potentially unstable gradients associated with the rendering process. The optimization is guided by a composite objective function that includes region-aware appearance losses for both the object and background, as well as a Dice-based mask loss for explicit geometric alignment.

\textbf{Problem Formulation.} Let $\mathbf{P}_c = (X_c, Y_c, Z_c) \in \mathbb{R}^3$ denote the position of the camera (or object) within the world coordinate system. Given an initial estimate $\mathbf{P}_{c}^{\text{init}}$, we define a bounded search space $\mathcal{B}$ for the coarse global sampling step as follows:
\begin{equation}
\mathcal{B} = [x_0 - \Delta_x, x_0 + \Delta_x] \times [y_0 - \Delta_y, y_0 + \Delta_y] \times [z_0 - \Delta_z, z_0 + \Delta_z],
\end{equation}
where $(x_0, y_0, z_0)$ corresponds to $\mathbf{P}_{c}^{\text{init}}$. For any candidate position $\mathbf{P} \in \mathcal{B}$, we generate a rendered image $I(\mathbf{P})$ and its associated object mask $M(\mathbf{P})$. The goal is to minimize the discrepancy between these outputs and the target image $I^{\ast}$ through a set of region-aware constraints.

\textbf{Region-Aware Appearance Loss.} To capture local photometric details, we define an appearance loss for the object region using the target mask $M^{\ast}$:
\begin{equation}
\mathcal{L}_{\text{obj}}(\mathbf{P}) = \frac{ \| (I(\mathbf{P}) - I^{\ast}) \odot M^{\ast} \|_1 }{ \| M^{\ast} \|_1 + \epsilon },
\end{equation}
where $\odot$ denotes element-wise multiplication and $\| \cdot \|_1$ represents the $L_1$ norm. The normalization term is critical to prevent the optimization process from biasing toward larger foreground areas. Similarly, the loss for the background region is formulated as:
\begin{equation}
\mathcal{L}_{\text{bg}}(\mathbf{P}) = \frac{ \| (I(\mathbf{P}) - I^{\ast}) \odot (1 - M^{\ast}) \|_1 }{ \| 1 - M^{\ast} \|_1 + \epsilon }.
\end{equation}

\textbf{Mask Alignment Loss.} To explicitly enforce geometric consistency between the rendered silhouette and the target, we employ a Dice-based loss:
\begin{equation}
\mathcal{L}_{\text{mask}}(\mathbf{P}) = 1 - \frac{2 \| M(\mathbf{P}) \odot M^{\ast} \|_1 + \epsilon}{ \| M(\mathbf{P}) \|_1 + \| M^{\ast} \|_1 + \epsilon}.
\end{equation}
This formulation ensures robustness against scale variations and is particularly effective for small foreground objects.

\textbf{Overall Objective.} The final objective function used to guide the Powell optimization is a weighted combination of the aforementioned losses, balanced by hyperparameters $\lambda$:
\begin{equation}
\mathcal{L}(\mathbf{P}) = \lambda_{\text{obj}} \mathcal{L}_{\text{obj}}(\mathbf{P}) + \lambda_{\text{bg}} \mathcal{L}_{\text{bg}}(\mathbf{P}) + \lambda_{\text{mask}} \mathcal{L}_{\text{mask}}(\mathbf{P}).
\end{equation}



\begin{figure}[!t]
    \centering
    \includegraphics[width=\linewidth]{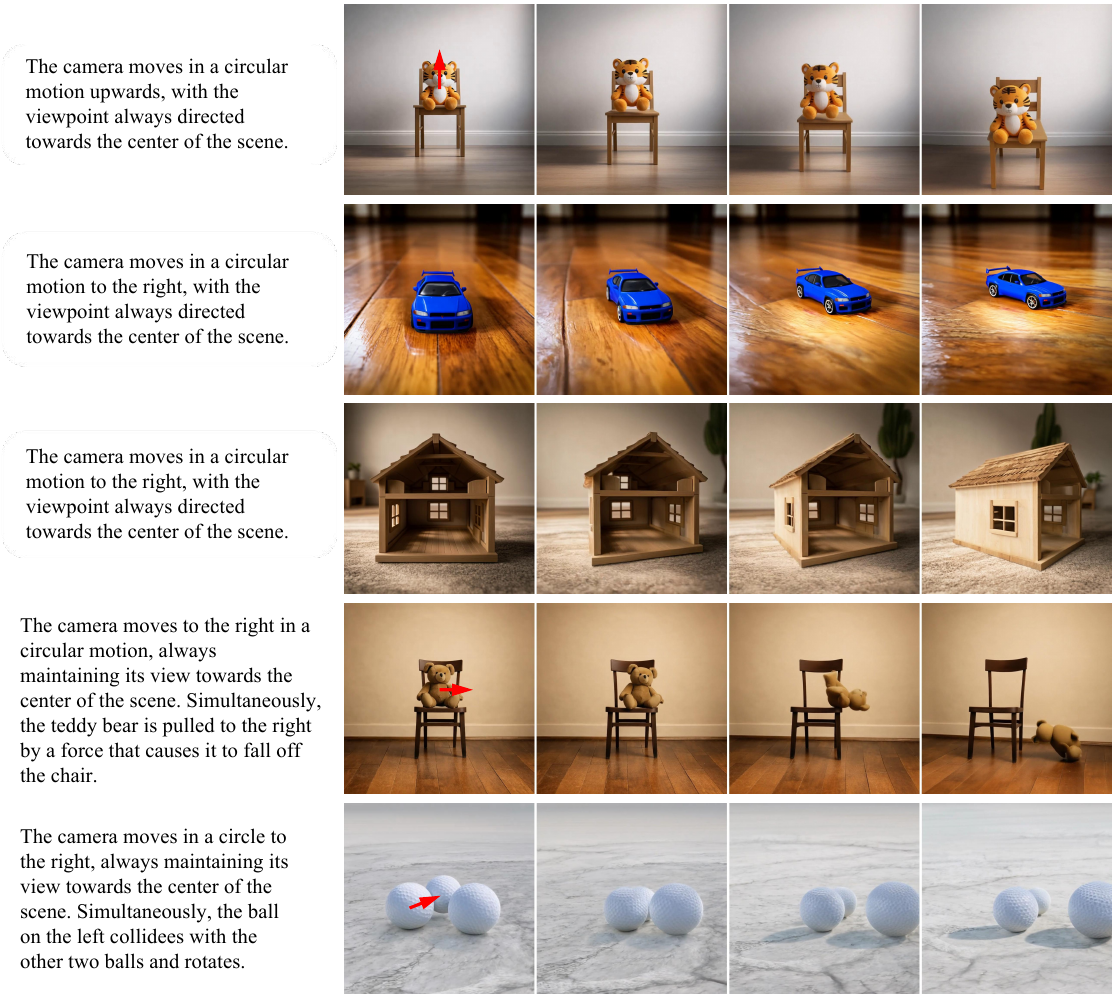}
    \caption{Illustration of camera motion and viewpoint variation}
    \Description{A diagram of a camera moving along a surrounding trajectory around the reconstructed scene, showing how the rendered viewpoint changes as the camera orbits the objects.}
    \label{fig:surround_diagram}
\end{figure}

\section{Details of Physical Simulation}
\label{sec:appendix_physics}

\subsection{VLM-Based Physics Configuration}
\label{app:vlm_prompt}

We use Qwen2.5-VL-72B-Instruct~\citep{Qwen2VL} to automatically estimate per-object physics materials and scene force fields from a single image. The VLM receives: (1) the full scene image, and (2) per-object crops extracted via segmentation masks. The complete system prompt is shown in Figure~\ref{fig:vlm_prompt}.

\begin{figure}[!t]
\begin{tcolorbox}[colback=gray!5, colframe=gray!60, title=VLM System Prompt for Physics Configuration, fonttitle=\small\bfseries, fontupper=\scriptsize, boxrule=0.5pt, arc=2pt]
\texttt{You are an expert physics material analyst and simulation designer. I will show you a scene image followed by cropped images of individual objects.}

\vspace{2pt}
\texttt{\textbf{Task A: Object Materials}}\\
\texttt{For each object (numbered starting from 0), identify:}\\
\texttt{1. What the object is (e.g., sand castle, rubber duck)}\\
\texttt{2. Best-matching physics material type for INTERESTING DEFORMABLE simulation.}\\
\texttt{\quad IMPORTANT: Prefer non-rigid materials --- choose the MOST DEFORMABLE plausible interpretation.}\\
\texttt{\quad Available material types:}

\vspace{2pt}
\texttt{\quad \textit{MPM materials} (particle-based, fluids/deformation):}\\
\texttt{\quad - "mpm\_liquid": liquids, viscous fluids. Params: E, nu, rho, viscous}\\
\texttt{\quad - "mpm\_elastoplastic": permanent deformation (clay, cream). Params: E, nu, rho, use\_von\_mises, yield\_stress}\\
\texttt{\quad - "mpm\_elastic": elastic deformable (rubber, jelly). Params: E, nu, rho, model}\\
\texttt{\quad - "mpm\_sand": granular (sand, powder). Params: E, nu, rho, friction\_angle}\\
\texttt{\quad - "mpm\_snow": snow/ice. Params: E, nu, rho, yield\_lower, yield\_higher}

\vspace{2pt}
\texttt{\quad \textit{PBD materials} (position-based dynamics):}\\
\texttt{\quad - "pbd\_cloth": thin sheet (fabric, paper). Params: rho, friction, compliance, air\_resistance}\\
\texttt{\quad - "pbd\_elastic": 3D soft body (sponge). Params: rho, friction, compliance}\\
\texttt{\quad - "pbd\_liquid": position-based fluid. Params: rho, density/viscosity relaxation}

\vspace{2pt}
\texttt{\quad \textit{Rigid} (use sparingly, ONLY for immovable structures):}\\
\texttt{\quad - "rigid": ground, wall, table. Do NOT use for small or deformable objects.}

\vspace{2pt}
\texttt{3. material\_params: E (Young modulus), rho (density), nu (Poisson ratio)}\\
\texttt{4. fixed: true ONLY if truly static}\\
\texttt{5. surface\_color: RGB float [0--1]}

\vspace{4pt}
\texttt{\textbf{Task B: Force Fields}}\\
\texttt{Suggest 1--3 force fields for interesting dynamics. Types:}\\
\texttt{constant, wind, point, drag, turbulence, vortex}\\
\texttt{Each with: direction, strength, start\_frame (-1 = immediate).}

\vspace{2pt}
\texttt{Respond with ONLY a JSON object: \{"objects": [...], "forces": [...]\}}
\end{tcolorbox}
\caption{Complete VLM prompt for automatic physics configuration. The prompt instructs Qwen2.5-VL-72B to identify per-object materials and suggest force fields from the scene image and per-object crops.}
\label{fig:vlm_prompt}
\Description{A text box containing the full vision-language-model prompt used for physics configuration, which asks the model to return a JSON object listing each object's material and suggested force fields based on the scene image and per-object crops.}
\end{figure}

\paragraph{Parameter Guidance.}
We provide the VLM with parameter ranges to ensure physically plausible outputs:
\begin{itemize}[nosep,leftmargin=*]
    \item Young's modulus $E$: foam $\sim 10^4$, jelly $\sim 10^3$, rubber $\sim 10^6$, glass $\sim 10^5$
    \item Density $\rho$ ($\text{kg/m}^3$): foam $\sim 50$, cream $\sim 500$, rubber $\sim 1100$, clay $\sim 1800$, glass $\sim 2500$
    \item Poisson's ratio $\nu$: typical $0.2$--$0.45$; nearly incompressible $\approx 0.45$
\end{itemize}

\paragraph{Postprocessing.}
The VLM output JSON is parsed with regex-based extraction. The parser accepts both the object form \texttt{\{"objects":...\}} and legacy array formats. Unknown material types fall back to \texttt{mpm\_elastic}. Invalid force types are discarded. All parameters are validated against type-specific defaults (Table~\ref{tab:material_defaults}).

Table~\ref{tab:material_defaults} lists all supported physics materials and their default parameters. Table~\ref{tab:force_fields} lists the available force field types.

\begin{table}[htb]
\caption{Supported physics material types and default parameters.}
\label{tab:material_defaults}
\begin{center}
\begin{adjustbox}{max width=\linewidth,center}
\begin{tabular}{llp{9cm}}
\toprule
\textbf{Material} & \textbf{Solver} & \textbf{Default Parameters} \\
\midrule
\texttt{rigid} & Rigid Body & $\rho = 200$, $\mu = 0.7$ \\
\texttt{mpm\_elastic} & MPM & $E = 3 \times 10^5$, $\nu = 0.2$, $\rho = 1000$, $\text{model} = \text{corotation}$ \\
\texttt{mpm\_elastoplastic} & MPM & $E = 3 \times 10^4$, $\nu = 0.4$, $\rho = 100$, von Mises yield = $10^4$ \\
\texttt{mpm\_sand} & MPM & $E = 5 \times 10^5$, $\nu = 0.2$, $\rho = 1800$, friction angle = $45^\circ$ \\
\texttt{mpm\_liquid} & MPM & $E = 10^6$, $\nu = 0.2$, $\rho = 1000$, viscous = false \\
\texttt{mpm\_snow} & MPM & $E = 10^6$, $\nu = 0.2$, $\rho = 1000$, yield $\in [0.025, 0.0045]$ \\
\texttt{mpm\_muscle} & MPM & $E = 10^6$, $\nu = 0.2$, $\rho = 1000$, $\text{model} = \text{Neo-Hookean}$ \\
\texttt{pbd\_elastic} & PBD & $\rho = 1000$, stretch/bending/volume compliance = 0, relaxation = 0.1 \\
\texttt{pbd\_cloth} & PBD & $\rho = 4\,\text{kg/m}^2$, stretch compliance = $10^{-7}$, bending = $10^{-5}$, air resistance = $10^{-3}$ \\
\texttt{pbd\_liquid} & PBD & $\rho = 1000$, density relaxation = 0.2, viscosity relaxation = 0.01 \\
\texttt{pbd\_particle} & PBD & $\rho = 1000$ \\
\bottomrule
\end{tabular}
\end{adjustbox}
\end{center}
\end{table}

\begin{table}[htb]
\caption{Supported force field types and default parameters.}
\label{tab:force_fields}
\begin{center}
\begin{adjustbox}{max width=\linewidth,center}
\begin{tabular}{lp{6cm}}
\toprule
\textbf{Type} & \textbf{Default Parameters} \\
\midrule
\texttt{constant} & direction $= [0,0,-1]$, strength $= 9.8$ \\
\texttt{wind} & direction $= [1,0,0]$, strength $= 1.0$, radius $= 1.0$ \\
\texttt{point} & strength $= 1.0$, position $= [0,0,0]$, falloff power $= 0$ \\
\texttt{drag} & linear $= 0$, quadratic $= 0$ \\
\texttt{vortex} & direction $= [0,0,1]$, perpendicular strength $= 20.0$ \\
\texttt{turbulence} & strength $= 1.0$, frequency $= 3$ \\
\texttt{noise} & strength $= 1.0$ \\
\bottomrule
\end{tabular}
\end{adjustbox}
\end{center}
\end{table}

\subsection{Multi-Physics Solvers Formulation}
\label{app:solvers}

As discussed in the Physical Simulation section of the main paper, our system relies on a unified multi-physics backend, Genesis~\cite{Genesis}, which integrates three specialized solvers. The detailed formulations and integration strategies are described below.

\textbf{Rigid Body Dynamics (RBD).} We utilize an RBD solver based on articulated body algorithms~\cite{featherstone2014rigid} for non-deformable objects and robotic manipulators. This method ensures stable, penetration-free interactions and accurate friction handling. The dynamics are governed by the generalized equation of motion:
\begin{equation}
    \mathbf{M}(\mathbf{q})\ddot{\mathbf{q}} + \mathbf{C}(\mathbf{q}, \dot{\mathbf{q}}) = \boldsymbol{\tau} + \mathbf{J}(\mathbf{q})^T \mathbf{f}_{\text{ext}},
\end{equation}
where $\mathbf{q}$ and $\dot{\mathbf{q}}$ denote the generalized coordinates and velocities, respectively. $\mathbf{M}$ represents the inertia matrix, $\mathbf{C}$ accounts for Coriolis and centrifugal forces, $\boldsymbol{\tau}$ denotes actuation torques, and $\mathbf{f}_{\text{ext}}$ represents external forces (e.g., contact) mapped into the joint space via the Jacobian $\mathbf{J}$. By solving for the acceleration $\ddot{\mathbf{q}}$, the solver updates the kinematic state of rigid entities without the numerical dissipation typical of particle-based methods.

\textbf{Material Point Method (MPM).} We employ MPM~\cite{sulsky1994mpm,hu2018moving,jiang2016mpm,klar2016drucker,ram2015viscoelastic,stomakhin2013material} to simulate fluids and hyperelastic materials. As a hybrid Lagrangian-Eulerian method, MPM tracks mass and momentum on Lagrangian particles ($p$) while computing forces on a background Eulerian grid ($i$). The governing equation for the grid node force $\mathbf{f}_i$ is derived from the divergence of particle stress $\boldsymbol{\sigma}_p$:
\begin{equation}
    \mathbf{f}_i = \sum_{p} V_p \boldsymbol{\sigma}_p \nabla w_{ip} + \mathbf{f}_{\text{ext}},
\end{equation}
where $V_p$ is the particle volume and $\nabla w_{ip}$ is the gradient of the interpolation kernel function connecting particle $p$ to grid node $i$. In each time step, particle states are transferred to the grid to solve the momentum equation, and the updated grid velocities are interpolated back to deform the particles, naturally facilitating the simulation of fracture, splashing, and plastic flow.

\textbf{Position-Based Dynamics (PBD).} We apply PBD~\cite{muller2007pbd} for thin-shell structures such as cloth and cables. Unlike force-based solvers that require small time steps for stability in stiff systems, PBD directly modifies object positions to satisfy geometric constraints. The position correction $\Delta \mathbf{x}_i$ for a particle $i$ is computed to satisfy a constraint function $C(\mathbf{x}) = 0$:
\begin{equation}
    \Delta \mathbf{x}_i = - w_i \frac{C(\mathbf{x})}{\sum_j w_j |\nabla_{\mathbf{x}_j} C(\mathbf{x})|^2} \nabla_{\mathbf{x}_i} C(\mathbf{x}),
\end{equation}
where $w_i$ is the inverse mass of particle $i$. By iteratively projecting positions onto the constraint manifold, PBD ensures that cloth remains inextensible and drapes naturally.

\textbf{Solver Integration and Optimization.} The integration of these three solvers is critical for high-fidelity multi-material simulations. The solvers operate concurrently, with interactions managed by transferring forces and state information at each simulation step. For instance, contact forces from the RBD solver are propagated to the MPM solver to induce material deformation, while the PBD solver enforces constraints during interactions with rigid bodies. To address computational challenges, we leverage GPU-based parallelism. Each solver operates independently on its respective domain subset, utilizing spatial partitioning to optimize interaction computations. Furthermore, adaptive time-stepping dynamically adjusts the simulation frequency based on local material properties, ensuring a balance between accuracy and efficiency.

\section{Details of WonderTrace}
\label{sec:appendix_video_details}

This section provides comprehensive details regarding the implementation of our video generation module, which consists of the WonderTrace rerendering pipeline and the scene construction strategy introduced in the main paper.

\textbf{WonderTrace Rerendering Details.} As highlighted in the main paper, videos rendered directly from physical simulations often exhibit a synthetic appearance. This visual deficiency stems from simplified lighting models, unrealistic material responses, and the absence of environmental noise. Consequently, such videos display a ``plastic'' texture that fails to capture the stochastic richness of real-world scenes, such as subtle atmospheric scattering and organic texture variations. To address this limitation, we employ video generation models to rerender simulation outputs, utilizing the physical simulation as a strict structural prior. By leveraging the high-dimensional latent space of state-of-the-art video generative models, we perform pixel-level semantic enhancement to synthesize photorealistic details onto the simulated structure.

To balance computational efficiency with visual fidelity, our implementation utilizes a partial denoising strategy. Rather than generating videos from pure Gaussian noise---a process that is computationally intensive and prone to deviation from the underlying simulation logic---we perturb the simulation frames with a moderate level of noise and apply only the final denoising stages of the diffusion process. 

Specifically, our primary pipeline adopts the Wan2.1 VACE~\cite{wan2025wan} model, which incorporates a self-forcing DMD-based distillation strategy. This design enables the rerendering process to operate with a minimal number of refinement steps, facilitating real-time performance without compromising visual quality. This approach significantly reduces inference latency while ensuring that the generated video faithfully inherits the complex motion trajectories and spatiotemporal structure of the simulation. During these final refinement stages, the model enriches the scene with high-frequency visual details---such as atmospheric scattering, sub-surface reflections, and motion blur---that are typically challenging for real-time physical simulators to model accurately.

Furthermore, we provide an alternative rendering configuration based on the larger Wan2.2 VACE 14B~\cite{wan2025wan} model. This variant is designed to maximize perceptual realism, delivering higher-resolution outputs with improved visual quality. It is particularly suitable for offline rendering or high-fidelity visualization scenarios where real-time constraints are relaxed.

\textbf{Scene Construction and Expansion.} A fundamental challenge in 3D scene modeling is the inherent sparsity of source data, which typically captures a limited field of view (FOV). Conventional reconstruction pipelines often fail to maintain consistent visual fidelity across unconstrained viewing angles; the absence of wide-field contextual imagery results in voids or severe boundary artifacts when the virtual camera trajectory deviates from the original path. This restricted FOV constrains the synthesis of immersive environments and limits the degrees of freedom for virtual camera movement.

Following the background handling approach described in the main paper, we employ a combined in-painting and out-painting strategy to effectively decouple the foreground content from the environment. We extend this design to the video rerendering stage to enable scene completion well beyond the original camera frustum. As illustrated in Fig.~\ref{fig:surround_diagram}, the reconstructed background is spatially expanded to provide sufficient semantic context for extensive camera motions and severe viewpoint changes. Finally, the rerendered, expanded background is seamlessly composited with the dynamic foreground elements. This dual-stream processing ensures continuous visual fidelity and robust temporal consistency under entirely novel camera trajectories.

\section{Details of Experiments}

\subsection{Simulation Details}
\label{app:simulation}

\paragraph{Physics Engine.}
We build on Genesis~\citep{Genesis}, a GPU-accelerated differentiable physics engine.
The simulation operates at $\Delta t = 4\,\text{ms}$ with 10 substeps per step, using the MPM (Material Point Method) solver for continuum materials and PBD (Position-Based Dynamics) for cloth and particles.
Rigid--MPM and Rigid--PBD coupling is enabled via the legacy coupler.
The MPM domain spans $[-2, 2]^3$ with particle size $0.01$.

\paragraph{Rendering.}
Each simulation frame is composited via a shadow-aware pipeline:
\begin{enumerate}[nosep,leftmargin=*]
    \item Render with segmentation: obtain RGB, segmentation IDs per pixel.
    \item Extract object mask ($\text{seg\_id} \geq 2$) and plane shadow mask ($\text{seg\_id} = 1$ and brightness $< 0.3$).
    \item Composite: $F = I_{\text{bg}} \cdot (1 - \alpha_{\text{obj}}) + I_{\text{render}} \cdot \alpha_{\text{obj}}$, then apply shadow darkening with strength $0.3$.
\end{enumerate}
Resolution is fixed at $880 \times 880$.

\paragraph{PBD Cloth Fixation.}
For cloth-like objects (e.g., dresses), we support a \texttt{fix\_top\_ratio} parameter that pins the topmost $r$\% of particles by $z$-coordinate after scene building, simulating hanging or attachment points.

\paragraph{Camera Motion.}
Six camera motion modes are supported: four orbital (around XY or YZ axes, clockwise/counterclockwise), one lateral, and one descent.
The angular velocity is $v = 0.001$ rad/frame at $60$ FPS.

\paragraph{Runtime.}
Table~\ref{tab:speed} reports the runtime breakdown on a single NVIDIA H100 GPU.

\begin{table}[htbp]
\caption{Average runtime breakdown of \modelname{}.}
\label{tab:speed}
\centering
\begin{adjustbox}{max width=\linewidth,center}
\begin{tabular}{lc}
\toprule
\textbf{Component} & \textbf{Time / Throughput} \\
\midrule
\multicolumn{2}{l}{\textit{One-time processing (per scene, pure inference)}} \\
\quad Segmentation (SAM3 + inpainting) & $\sim$5 s \\
\quad 3D Mesh Reconstruction (SAM3D) & $\sim$15 s \\
\quad Physics Config Estimation (VLM) & $\sim$10 s \\
\quad Scene Alignment (pose + camera opt.) & $\sim$16 s \\
\midrule
\multicolumn{2}{l}{\textit{Interactive simulation loop}} \\
\quad End-to-end Interactive Loop & $\sim$15 FPS \\
\bottomrule
\end{tabular}
\end{adjustbox}
\parbox{\linewidth}{\footnotesize Processing timings represent highly optimized pure model inference, excluding model loading overhead. Physics configuration estimation uses Qwen2.5-VL-72B for automatic material and force prediction. The interactive loop throughput corresponds to a real-time interactive physics preview. Optional high-fidelity offline diffusion rerendering can be applied afterward for enhanced photorealism.}
\end{table}

\subsection{Mesh Reconstruction Details}
\label{app:mesh_recon}

\paragraph{Segmentation.}
We use SAM3 for text-prompted instance segmentation.
Given text prompts (e.g., ``glass ball'', ``sand castle''), SAM3 produces per-object binary masks $\{M_k\}$.
A combined binary mask $\bigcup_k M_k$ is computed for background inpainting.

\paragraph{Background Inpainting.}
LaMa inpaints the masked regions to produce a clean background $I_{\text{bg}}$.
We apply cumulative inpainting: objects are removed sequentially to handle overlapping masks.
Dilation kernel size is $100$ pixels.

\paragraph{3D Reconstruction.}
SAM3D reconstructs per-object 3D meshes from single-view images with per-object masks.
The model predicts rotation (quaternion), scale, and translation for each object, which are composed into an affine transform:
\begin{equation}
    \mathbf{x}_{\text{world}} = s \cdot \mathbf{R}_q \, \mathbf{x}_{\text{local}} + \mathbf{t} ,
\end{equation}
where $\mathbf{R}_q$ is the rotation matrix from the predicted quaternion, $s$ is the scale factor, and $\mathbf{t}$ is the translation.
A coordinate frame conversion (Y-up to Z-up) is applied before exporting as OBJ files.

\subsection{Evaluation Metrics}
\label{app:metrics}

\paragraph{Image Quality.}
\begin{itemize}[nosep,leftmargin=*]
    \item \textbf{SSIM}: Structural Similarity~\citep{wang2004image} between rendered first frame and input image (grayscale, data range 255).
    \item \textbf{LPIPS}: Learned Perceptual Image Patch Similarity~\citep{zhang2018unreasonable} using AlexNet backbone. Falls back to normalized MSE ($\min(4 \cdot \text{MSE}, 1)$) when model weights are unavailable.
\end{itemize}

\paragraph{Alignment Metrics.}
\begin{itemize}[nosep,leftmargin=*]
    \item \textbf{Mask IoU}: Binary intersection-over-union between the rendered object silhouette and the GT segmentation mask.
    \item \textbf{Reproj.\ Error}: $L_2$ distance (pixels) between the centroids of the predicted and GT masks.
\end{itemize}

\paragraph{Physics Metrics.}
\begin{itemize}[nosep,leftmargin=*]
    \item \textbf{Penetration Rate (PR)}: Fraction of object pairs with AABB overlap. Lower is better.
    \item \textbf{Support Violation Rate (SVR)}: Fraction of objects with $z_{\min} > 0.02$ (floating above ground). Lower is better.
    \item \textbf{Interaction Success Rate (ISR)}: Fraction of frames where the object mask covers $> 0.1\%$ of the image area ($880 \times 880$). Higher is better.
\end{itemize}

\paragraph{Video Quality.}
\begin{itemize}[nosep,leftmargin=*]
    \item \textbf{Motion Amplitude}: Mean optical flow magnitude (Farneback) across consecutive frames.
    \item \textbf{Motion Smoothness}: Inverse of frame-to-frame flow variance: $S = 1 / (1 + \text{Var}(\{\bar{f}_t\}))$.
    \item \textbf{Aesthetic}: CLIP-based (ViT-B/32) cosine similarity with ``a beautiful high quality scene''. Falls back to sigmoid-normalized Laplacian variance: $A = 2 / (1 + e^{-\bar{\sigma}^2_L / 500}) - 1$.
\end{itemize}

\section{Quantitative Evaluation via GPT-5}
\label{sec:appendix_gpt_eval}

To quantitatively evaluate the 60-scene test set for controllability, physical plausibility, and overall video quality, we adopt an automated Vision-Language Model (VLM) scoring framework inspired by the VideoPhy~\cite{bansal2024videophy} protocol. Specifically, we utilize a GPT-5-based evaluator to assign 5-point Likert scores to the generated videos. 

\subsection{Data Preparation and Frame Sampling}
To construct the evaluation queries while adhering to the context limits of the VLM, we preprocess the input images and generated videos as follows:
\begin{itemize}
    \item \textbf{Frame Extraction:} For each generated video, we uniformly sample 10 evenly spaced frames across the temporal axis to represent the full sequence. 
    \item \textbf{Resolution Scaling:} Both the initial input image and the sampled video frames are downscaled such that their longest edge is 880 pixels, preserving the aspect ratio. 
    \item \textbf{Visual Prompts:} The input image explicitly visualizes the initial motion position and direction using a red arrow, providing the model with a clear spatial reference for the applied physical constraints. All images and frames are converted into base64 JPEG format before being parsed to the VLM.
\end{itemize}

\subsection{Evaluation Prompt and Criteria}
The VLM is provided with the text prompt, the initial condition image, and the sequences of 10 sampled frames from the compared models (presented in chronological order). The model is instructed to assess the videos on a scale of 1 (poor) to 5 (excellent) based on three complementary criteria: Semantic Adherence, Physical Commonsense, and Video Quality. 

The exact prompt template used for the evaluation is provided below:

\begin{quote}
\textit{You are tasked with evaluating the quality of image-to-video generation produced by a model. For each test case, you will be given:}\\
\textit{1. A text prompt describes one or more objects along with the initial motion direction. The motion's position and direction are visualized as a red arrow in the input image.}\\
\textit{2. An input image of the object.}\\
\textit{3. Eight sets of 10 evenly spaced frames—each set corresponds to a video generated by a different model from the same input.}\\

\textit{Please evaluate this video based on the following three criteria using a 5-point Likert scale (1 = poor, 5 = excellent):}\\
\textit{- \textbf{Semantic Adherence:} How well the content and motion in the video match the description in the text prompt, especially the alignment with the motion direction and position. Note that the video should starts with the input image.}\\
\textit{- \textbf{Physical Commonsense:} Whether the object’s motion follows intuitive, physically plausible dynamics given the applied force direction and position.}\\
\textit{- \textbf{Video Quality:} The overall visual and temporal quality of the video (note that static or nearly-static sequences are less preferred).}\\

\textit{Provide your evaluation for each video strictly in the following one-line format:}\\
\textit{Model i, Semantic Adherence score, Physical Commonsense score, Video Quality score}
\end{quote}

\subsection{GPT-Free Evaluation on VBench-2.0}
\label{app:vbench2}

The GPT-5-based protocol above follows the widely used VideoPhy~\cite{bansal2024videophy} VLM-as-judge paradigm and is not a metric introduced by this work. To further remove any dependence on a proprietary judge, we additionally report an open-source, GPT-free evaluation using VBench-2.0. Table~\ref{tab:vbench2} reports four physics-related dimensions---rigidity (Rigid), non-penetration (NoPen), object permanence (Perm), and motion coherence (MotCoh)---together with their average. \modelname{} attains the highest average physics score and leads on rigidity and permanence, corroborating the GPT-5-based results with a fully reproducible, open-source metric.

\begin{table}[t]
\centering
\caption{GPT-free evaluation on VBench-2.0. We report four physics-related dimensions and their average. Rigid: rigidity; NoPen: non-penetration; Perm: object permanence; MotCoh: motion coherence.}
\label{tab:vbench2}
\begin{adjustbox}{max width=\linewidth,center}
\renewcommand{\arraystretch}{0.95}
\begin{tabular}{lccccc}
\toprule
\textbf{Method} & \textbf{Rigid} & \textbf{NoPen} & \textbf{Perm} & \textbf{MotCoh} & \textbf{Avg} \\
\midrule
\textbf{Ours} & \textbf{0.98} & 0.84 & \textbf{1.00} & 0.98 & \textbf{0.95} \\
Veo3.1 & \textbf{0.98} & 0.83 & 0.98 & 0.98 & 0.94 \\
Wan2.2-A14B & 0.95 & 0.82 & \textbf{1.00} & 0.98 & 0.94 \\
Sora2-pro & \textbf{0.98} & \textbf{0.86} & 0.98 & 0.93 & 0.94 \\
CogVideoX1.5 & 0.77 & 0.82 & \textbf{1.00} & \textbf{1.00} & 0.90 \\
Wan2.2-5B & 0.93 & 0.81 & 0.95 & 0.91 & 0.90 \\
PhysCtrl & 0.80 & 0.78 & \textbf{1.00} & 0.97 & 0.89 \\
\bottomrule
\end{tabular}%
\end{adjustbox}
\end{table}

\subsection{Hyperparameter Settings}
\label{app:hyperparams}

Table~\ref{tab:hyperparams} lists the key hyperparameters and their values used throughout the experiments.

\begin{table}[t]
\caption{Key hyperparameters.}
\label{tab:hyperparams}
\begin{center}
\begin{tabular}{llc}
\toprule
\textbf{Component} & \textbf{Parameter} & \textbf{Value} \\
\midrule
\multirow{4}{*}{Physics} & Time step $\Delta t$ & $4\,\text{ms}$ \\
 & Substeps & 10 \\
 & Default simulation steps $T$ & 300 \\
 & Render FPS & 60 \\
\midrule
\multirow{4}{*}{RANSAC} & Distance threshold $\tau_d$ & 0.01 \\
 & Min.\ samples $n_{\text{ransac}}$ & 3 \\
 & Iterations & 2{,}000 \\
 & Horizontality threshold $\tau_{\cos}$ & 0.8 \\
\midrule
\multirow{4}{*}{Camera Opt.} & Random search samples $N_{\text{rand}}$ & 60 \\
 & Powell max iterations & 80 \\
 & Object loss weight $w_{\text{obj}}$ & 1.0 \\
 & Background loss weight $w_{\text{bg}}$ & 0.2 \\
 & Mask loss weight $w_{\text{mask}}$ & 1.0 \\
\midrule
\multirow{3}{*}{Video Synthesis} & Denoising steps & 10 \\
 & Number of frames & 81 \\
 & Output FPS & 15 \\
\midrule
\multirow{2}{*}{Penetration} & AABB padding & 0.01 \\
 & Max displacement $\delta_{\max}$ & 0.05 \\
\bottomrule
\end{tabular}
\end{center}
\end{table}

\section{Additional Ablation Study}
\label{appendix:ablation_study}
\begin{table}[htbp]
\caption{Ablation on force field configuration.}
\label{tab:ablation_forces}
\centering
\begin{adjustbox}{max width=\linewidth,center}
\begin{tabular}{lccc}
\toprule
\textbf{Variant} & \textbf{Motion Amp.$\uparrow$} & \textbf{Smoothness$\uparrow$} & \textbf{Aesthetic$\uparrow$} \\
\midrule
No forces & \textbf{4.77} & \textbf{0.45} & \textbf{0.22} \\
Gravity only & 4.73 & 0.41 & 0.17 \\
Primary force only & 4.35 & 0.40 & 0.17 \\
Full forces (ours) & 4.35 & 0.40 & 0.17 \\
\bottomrule
\end{tabular}
\end{adjustbox}
\vspace{0.1cm}
\parbox{\linewidth}{\footnotesize Motion Amp.: mean optical flow magnitude. Smoothness: inverse flow variance.}
\end{table}

\textbf{Force field configuration.} 
Table~\ref{tab:ablation_forces} investigates the impact of force fields during the physical simulation stage. Interestingly, the \emph{No forces} variant achieves the highest scores in automated 2D metrics such as Motion Amplitude (4.77) and Smoothness (0.45). However, we argue that these purely appearance-based metrics inherently favor unconstrained, linear, or ``floating'' motions. When our \emph{Full forces} (including gravity and intended primary forces) are introduced, the objects are subjected to strict physical laws---such as sudden deceleration due to collisions, friction, and resting contacts. While these physical constraints naturally reduce the naive optical flow smoothness and overall motion amplitude, they are fundamentally required to prevent objects from drifting unnaturally. The parity between \emph{Primary force only} and \emph{Full forces} further suggests that explicit external driving forces dominate the valid dynamic interactions in our targeted events.

\begin{table}[htbp]
\caption{Ablation on video synthesis inference steps.}
\label{tab:ablation_synthesis}
\centering
\begin{adjustbox}{max width=\linewidth,center}
\begin{tabular}{lcccc}
\toprule
\textbf{Variant} & \textbf{SSIM$\uparrow$} & \textbf{LPIPS$\downarrow$} & \textbf{Aesthetic$\uparrow$} & \textbf{Time (s)$\downarrow$} \\
\midrule
10 steps & \textbf{0.82} & \textbf{0.05} & \textbf{0.20} & \textbf{539.03} \\
30 steps & 0.79 & 0.06 & 0.16 & 1247.30 \\
50 steps & 0.79 & 0.06 & 0.15 & 1922.37 \\
\bottomrule
\end{tabular}
\end{adjustbox}
\vspace{0.1cm}
\parbox{\linewidth}{\footnotesize Time: wall-clock inference time per video in seconds.}
\end{table}

\textbf{Inference efficiency in video synthesis.} 
Finally, we study the trade-off between rendering quality and computational cost in the video synthesis stage (Table~\ref{tab:ablation_synthesis}). We observe that executing the diffusion process for \emph{10 steps} yields the optimal visual fidelity (SSIM $0.82$, LPIPS $0.05$, Aesthetic $0.20$) while requiring the least wall-clock time ($539.03$\,s). Increasing the inference to 30 or 50 steps exponentially increases the computational burden but slightly degrades the perceptual metrics. This indicates that our condition signals (rendered explicit dynamics) are highly informative, allowing the rerendering model to converge rapidly without requiring over-parameterized iterative refinement. Consequently, we adopt 10 steps as the default setting for optimal efficiency.

\textbf{Geometric Fidelity of Re-rendering.} We evaluate whether depth-conditioned video re-rendering (using 10 denoising steps) reliably preserves the underlying physical geometry of the raw simulation. As demonstrated in Tab.~\ref{tab:ablation_rerender}, the re-rendered outputs maintain robust scene layouts, achieving an overall SSIM of 0.72 and a background SSIM of 0.78. Fluid environments exhibit the highest structural fidelity (SSIM of 0.80, background SSIM of 0.84), indicating that depth maps provide strong geometric constraints for continuous surfaces. Additionally, deformable objects such as cloth demonstrate nearly zero silhouette drift ($\Delta\mathrm{IoU} \approx 0.00$). This structural consistency is further corroborated by an average centroid displacement of merely 52.5 pixels (6.0\% of the 880-pixel image width), confirming that precise spatial positioning is strictly maintained throughout the re-rendering process.

\begin{table}[t]
\centering
\caption{Geometric fidelity of video re-rendering. We measure structural similarity between raw physics simulation and depth-conditioned re-rendered output (10 denoising steps) to verify that the re-rendering preserves underlying physical geometry.}
\label{tab:ablation_rerender}
\begin{adjustbox}{max width=\linewidth,center}
\begin{tabular}{lccccc}
\toprule
\textbf{Material} & \textbf{SSIM$\uparrow$} & \textbf{BG-SSIM$\uparrow$} & \textbf{Obj-SSIM$\uparrow$} & \textbf{Centroid$\downarrow$} & \textbf{$\Delta$IoU} \\
\midrule
Fluid (7) & 0.80 & 0.84 & 0.56 & 42.3 & -0.34 \\
Rigid/Elastic (5) & 0.65 & 0.71 & 0.40 & 72.7 & -0.27 \\
Cloth (3) & 0.67 & 0.74 & 0.44 & 42.8 & -0.00 \\
\midrule
\textbf{Overall (15)} & \textbf{0.72} & \textbf{0.78} & \textbf{0.48} & \textbf{52.5} & \textbf{-0.25} \\
\bottomrule
\end{tabular}%
\end{adjustbox}

\vspace{1mm}
\raggedright
\footnotesize
SSIM: per-frame structural similarity (raw sim.~$\leftrightarrow$ re-rendered). BG/Obj-SSIM: SSIM within background/object regions. Centroid: mean object centroid displacement (px, image size 880$\times$880). $\Delta$IoU: mask IoU change vs.\ ground truth after re-rendering.
\end{table}

\section{Robustness and Failure Analysis}
\label{sec:appendix_failure}

\begin{figure}[p]
  \centering
  \includegraphics[width=\linewidth]{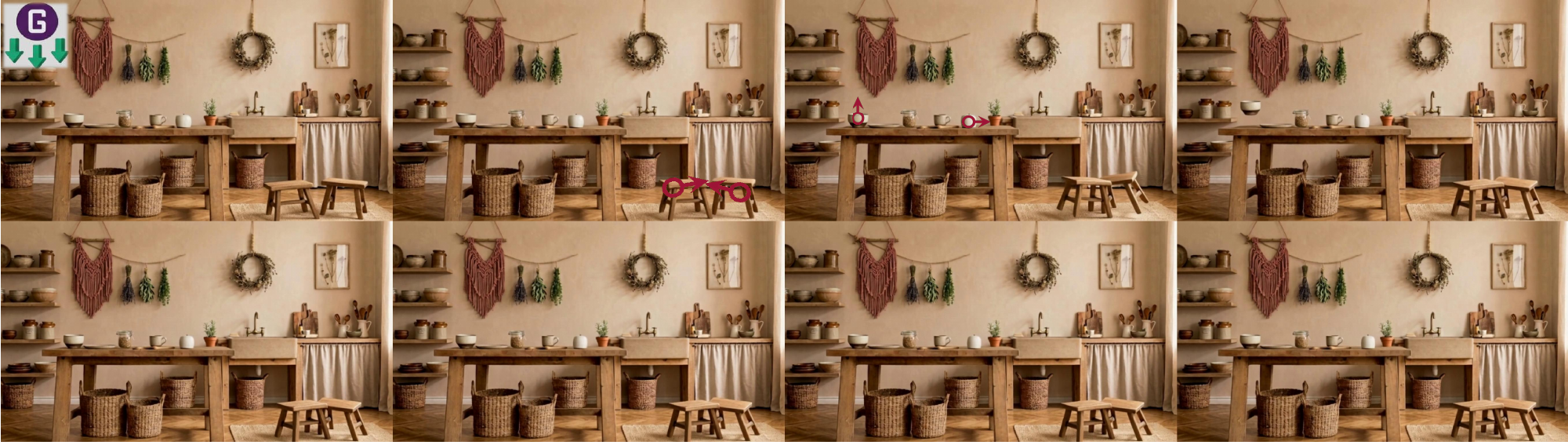}
  \caption{Robustness on a highly cluttered scene (a kitchen counter). Top-to-bottom, left-to-right shows the temporal evolution of the interactive physics preview. Despite dense clutter and heavy occlusion, \modelname{} reliably segments, reconstructs, aligns, and simulates the scene while remaining interactive.}
  \Description{A two-row, four-column sequence of frames of a cluttered kitchen-counter scene over time. Small arrows indicate applied forces on objects, and the objects respond with physically plausible motion while the dense surrounding clutter remains stable.}
  \label{fig:cluttered_kitchen}
\end{figure}

To quantify the robustness of \modelname{} and characterize its failure modes, we execute the full pipeline on 100 diverse scenes and manually inspect every intermediate and final result. Table~\ref{tab:failure_analysis} reports the per-stage failure counts and success rates. The dominant source of error is VLM-based material estimation (9 failures), followed by segmentation/decomposition and pose/scene alignment (4 each) and 3D lifting (3). These failures correspond to the challenges raised by reviewers---heavy occlusion, mis-segmentation, VLM material errors, and depth ambiguity---and remain relatively rare, yielding an overall usable rate of 0.89. We further stress-test the pipeline on highly cluttered kitchen-counter scenes (Fig.~\ref{fig:cluttered_kitchen}), where it reliably segments, reconstructs, and simulates the scene while remaining interactive.

\begin{table}[t]
\centering
\caption{Per-stage failure analysis on 100 scenes. We manually inspect each result and attribute failures to the earliest stage at which they occur.}
\label{tab:failure_analysis}
\begin{adjustbox}{max width=\linewidth,center}
\renewcommand{\arraystretch}{0.95}
\begin{tabular}{lcc}
\toprule
\textbf{Stage (100 scenes)} & \textbf{\# Failures} & \textbf{Success} \\
\midrule
Segmentation / decomposition (SAM3) & 4 & 0.96 \\
3D lifting (SAM-3D-Objects) & 3 & 0.97 \\
Pose / scene alignment & 4 & 0.96 \\
VLM material estimation & 9 & 0.91 \\
\midrule
\textbf{Overall usable results} & \textbf{11} & \textbf{0.89} \\
\bottomrule
\end{tabular}%
\end{adjustbox}
\end{table}

\FloatBarrier

\section{Additional Qualitative Results}
\label{sec:additional_results}

In this section, we provide further qualitative results to complement the main text. 

\begin{figure}[p]
  \centering
  \includegraphics[width=\linewidth]{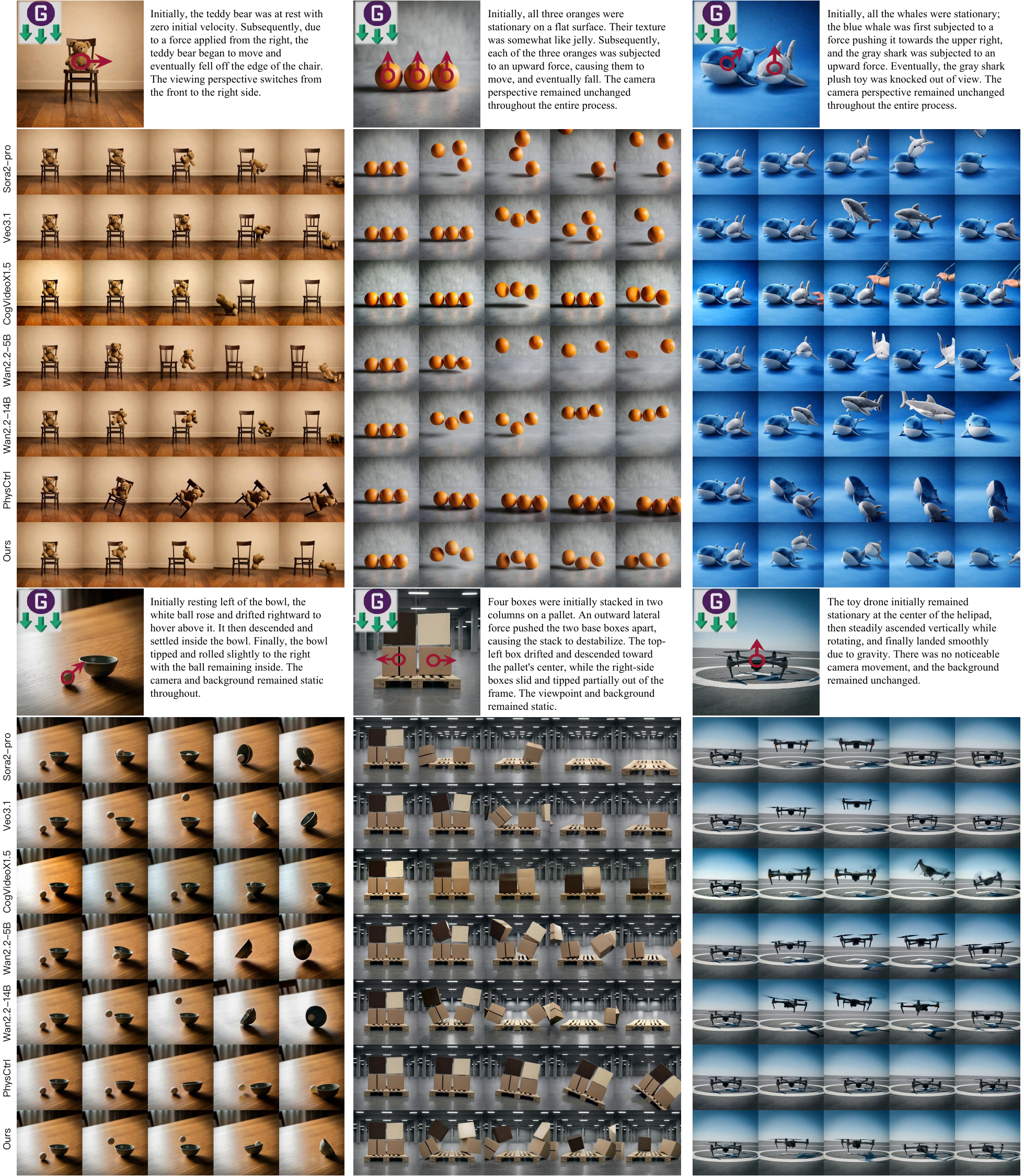}
  \caption{Qualitative comparison between the proposed method and existing video generation methods.}
  \Description{An extended comparison grid where each row is a test scene and each column is a method, showing additional qualitative examples of our method against existing video generation baselines across diverse scenes.}
  \label{fig:cpmpare}
\end{figure}

\begin{figure}[p]
  \centering
  \includegraphics[width=\linewidth]{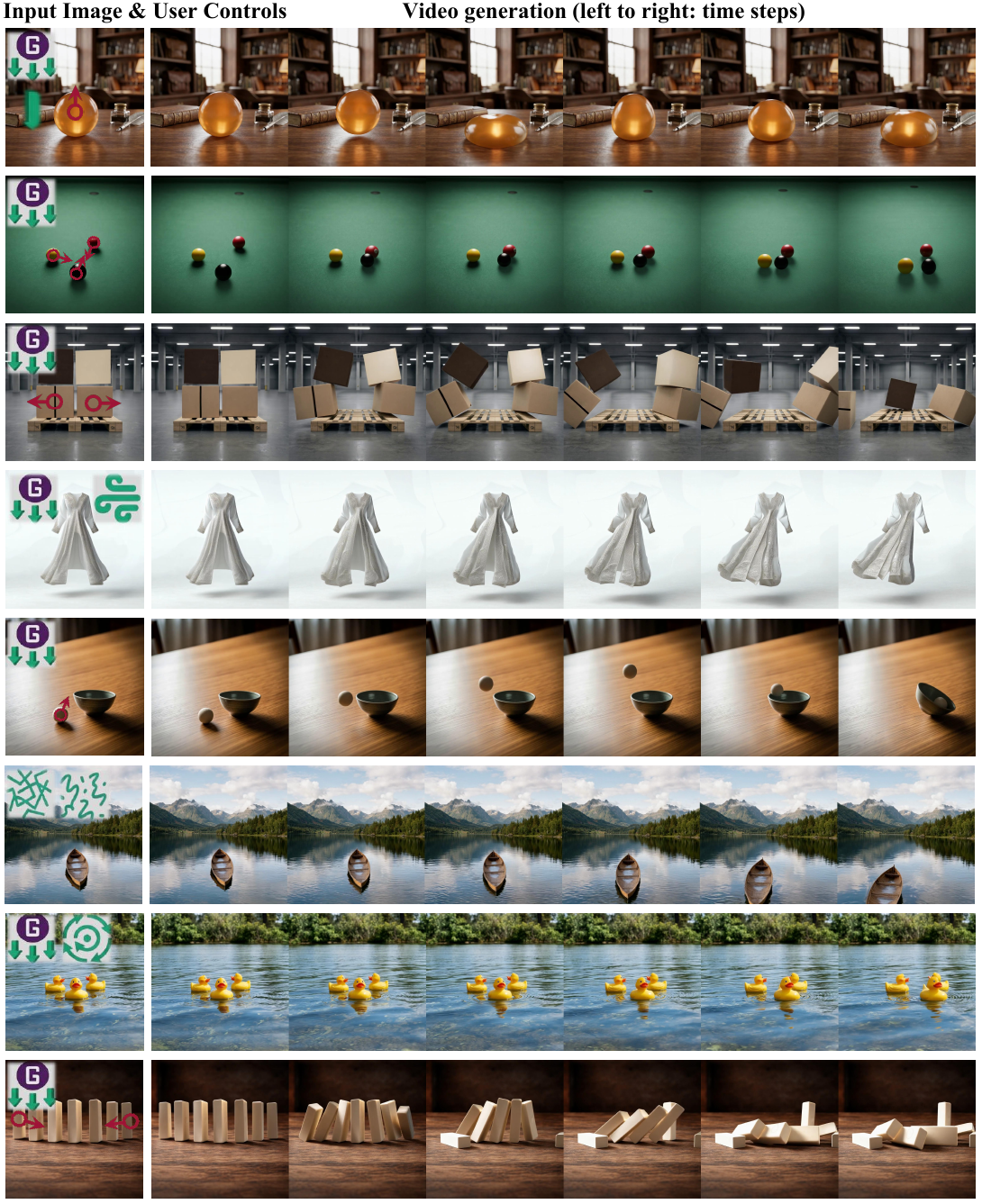}
  \caption{Qualitative results of the proposed method.}
  \Description{Additional visual results showing the model's performance on multi-object interactions.}
\end{figure}

\begin{figure}[p]
  \centering
  \includegraphics[width=\linewidth]{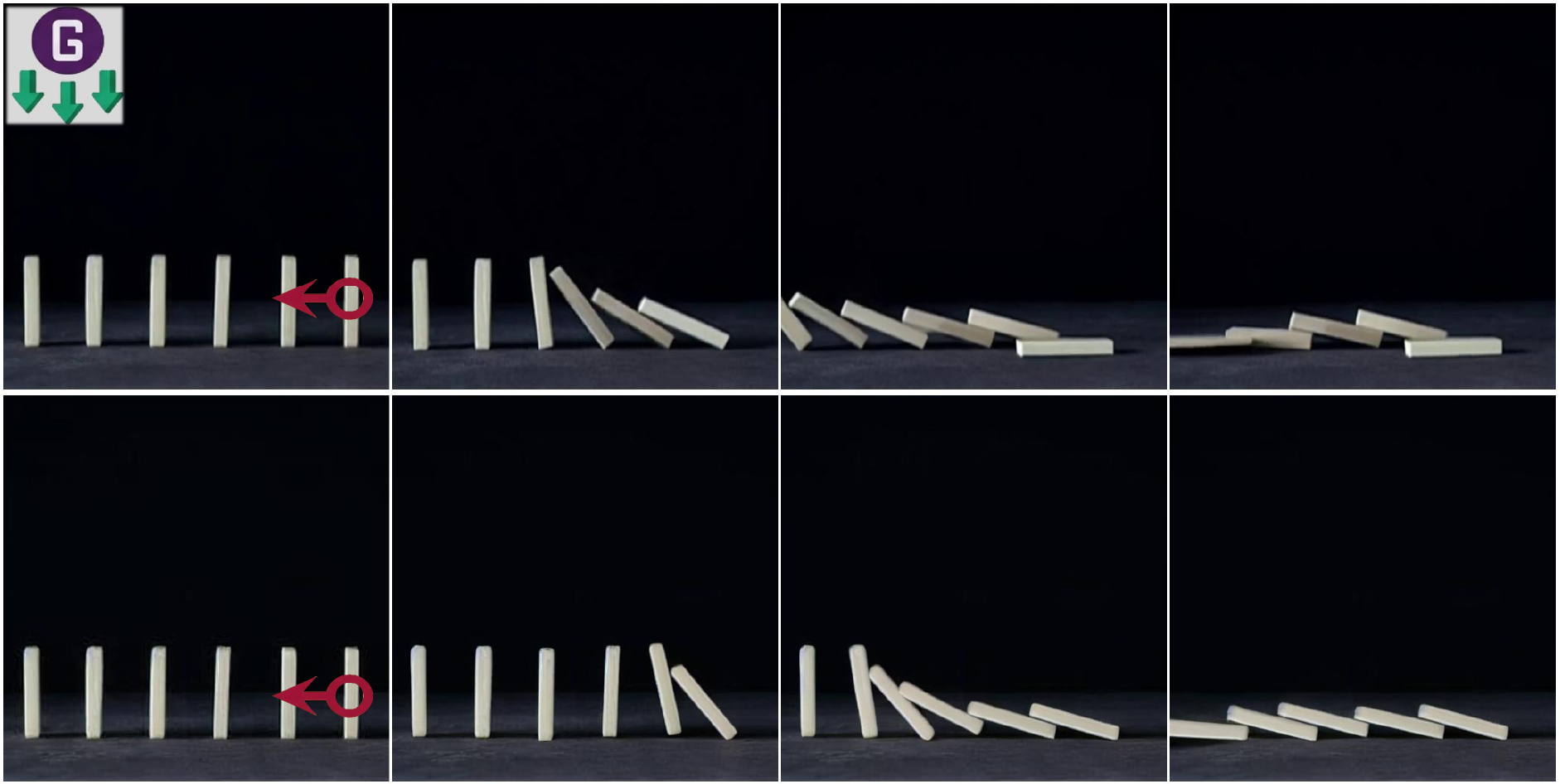}
  \caption{Direct qualitative comparison with PhysGen~\cite{liu2024physgen} on a domino-toppling event under the same applied force. Top: \modelname{} (Ours); bottom: PhysGen. \modelname{} yields a physically coherent sequential collapse, while PhysGen, operating in 2D image space, is less faithful to the underlying 3D dynamics.}
  \Description{Two rows of four frames each showing a row of domino blocks toppling after a leftward force is applied. The top row is our method and the bottom row is PhysGen; both start from the same standing configuration and progress to a fallen state, with our method showing more physically plausible sequential collapse.}
  \label{fig:physgen_compare}
\end{figure}

We further provide a direct comparison against PhysGen~\cite{liu2024physgen} in Fig.~\ref{fig:physgen_compare}. PhysGen targets 2D image-space manipulation and therefore cannot express the 3D force-field scene manipulation that \modelname{} performs; consequently, the stronger physics-aware baselines PhysCtrl and WonderPlay are used for the main comparisons. On a shared domino-toppling event, \modelname{} produces physically coherent sequential collapse consistent with the applied force, whereas PhysGen exhibits less faithful 3D dynamics.

\end{document}